\documentclass[11pt,a4paper]{article}

\usepackage{placeins}
\usepackage{graphicx}
\usepackage{xcolor}
\usepackage{float}
\usepackage{afterpage}
\usepackage{amssymb,amsmath,bm}
\usepackage{multirow,booktabs,ulem}
\usepackage{cite}
\usepackage{jheppub}
\usepackage{hyperref}
\usepackage{enumitem}
\usepackage{framed}

\usepackage{tikz}
\usetikzlibrary{shapes.geometric, arrows.meta, positioning}

\tikzstyle{roundednode} = [rectangle, rounded corners, minimum width=3cm, minimum height=1cm, text centered, draw=black, fill=blue!20]
\tikzstyle{arrow} = [thick, ->, >=Stealth]

\renewcommand{\vec}[1]{\textbf{#1}}

\usepackage{caption}
\usepackage{subcaption}

\newcommand*{\tmp}[4]{\ensuremath{%
    {#4%
    \ifx\empty#3\empty\ifx\empty#1\empty\else^{#1}\fi\else^{#1(#3)}\fi%
    \ifx\empty#2\empty\else_{#2}\fi}%
}}

\newcommand{\quotes}[1]{``#1''}

\usepackage{braket}

\title{A linear PDF model for Bayesian inference}
\author[a]{Mark N.~Costantini,}
\author[b]{Luca Mantani,}
\author[c]{James M.~Moore,}
\author[a]{Maria Ubiali}

\affiliation[a]{DAMTP, University of Cambridge, Wilberforce Road, Cambridge, CB3 0WA, United Kingdom}
\affiliation[b]{Instituto de F\'isica Corpuscular (IFIC), Universidad de Valencia-CSIC, E-46980 Valencia, Spain}
\affiliation[c]{Lucy Cavendish College, University of Cambridge, Lady Margaret Road, Cambridge, CB3 0BU, United Kingdom}
\emailAdd{mnc33@cam.ac.uk}
\emailAdd{luca.mantani@uv.es}
\emailAdd{jmm232@cam.ac.uk}
\emailAdd{M.Ubiali@damtp.cam.ac.uk}

\abstract{
A robust uncertainty estimate in global analyses of Parton Distribution Functions (PDFs) 
is essential at the Large Hadron Collider (LHC), especially in view of the high-precision 
data anticipated by experimentalists in the High-Luminosity phase of the LHC. 
A Bayesian framework to determine PDFs provides a rigorous treatment 
of uncertainties and full control on the prior, though its practical implementation can be computationally demanding.
To address these challenges, we introduce a novel approach to PDF determination tailored for Bayesian inference, based on the use of linear models. 
Unlike traditional parametrisations, our method represents PDFs as vectors in a functional space 
spanned by specially chosen bases, derived from the dimensional reduction of a neural network functional space, providing a compact yet versatile representation of PDFs.  
The low-dimensionality of the preferred models allows for particularly fast inference.  
The size of the bases can be systematically adjusted, allowing for transparent control over underfitting and overfitting, 
and facilitating principled model selection through Bayesian workflows.
In this work, the methodology is applied to a fit of Deep Inelastic Scattering synthetic data, and thoroughly tested via multi-closure tests,  
thus paving the way to its application to global PDF fits.

}

\begin{document}

\maketitle

\section{Introduction}
\label{sec:intro}
One of the key elements of any theoretical predictions at 
hadron colliders are the Parton Distribution Functions (PDFs) of the
proton; see Refs.~\cite{Amoroso:2022eow,Ubiali:2024pyg} for some recent reviews. 
At a time when precision studies at the Large Hadron Collider (LHC) represent the ultimate challenge for 
deepening our understanding of the Standard Model (SM) and identifying 
any possible signs of new physics, an accurate determination of these functions 
is crucial, particularly given the extreme accuracy of 
current LHC data and in light of the high-precision data 
anticipated from the High-Luminosity Large Hadron Collider (HL-LHC)~\cite{AbdulKhalek:2018rok}.  

The PDFs are functions of two variables, an energy scale $Q$, and a fraction of the proton's momentum $x$ carried by a parton. 
While the evolution of PDFs with the energy scale can be
computed using perturbative QCD, the dependence of these
functions on the fraction of proton's momentum $x$ is unknown, and currently must be determined from the available experimental data. 
The determination of continuous functions from a finite set of experimental 
observations constitutes an ill-posed inverse problem; its solution requires a regularisation procedure that reduces the infinite-dimensional 
functional space of PDFs to a finite-dimensional space, in which the inverse problem 
becomes solvable. 
Such a procedure unavoidably introduces some kind of bias, the extent of which depends on the 
specific methodology. A robust quantification of PDF uncertainties must take 
into account the methodological error associated with the regularisation procedure, 
alongside the other sources of experimental and theoretical errors~\cite{Ball:2018twp,NNPDF:2024dpb,Barontini:2025lnl}. 

In the fitting methodologies currently used for PDF determinations, the 
unknown PDF model is parametrised in terms of a finite (albeit rather large) set of 
parameters, which are then fitted to the experimental data.
Global PDF determinations widely used for phenomenological studies include
ABMP16~\cite{Alekhin:2017kpj}, CT18~\cite{Hou:2019efy}, MSHT20~\cite{Bailey:2020ooq,Cridge:2023ryv} and  
NNPDF4.0~\cite{NNPDF:2021njg,NNPDF:2024nan}. 
Significant effort has been invested into comparing the results obtained with different methodologies 
and assessing their performance on data that have not been included in any of the PDF analyses~\cite{Chiefa:2025loi}, 
as well as benchmarking the various methodologies and producing combinations of the available global PDF fits~\cite{Butterworth:2015oua,PDF4LHCWorkingGroup:2022cjn}.  
Moreover, a serious scrutiny of the robustness of the various approaches via statistical closure tests 
has been performed~\cite{DelDebbio:2021whr,Harland-Lang:2024kvt}, 
even in the presence of inconsistent experimental data~\cite{Barontini:2025lnl}.
Despite these crucial efforts increasing our understanding of the origin of the differences 
between PDF determinations, and highlighting some of the limits in the currently used methodologies, 
there is no evidence that a specific PDF set outperforms the others, or that 
one of the global PDF sets should not be used in phenomenological analyses at the LHC. 

At the same time, recent studies by the ATLAS and CMS collaborations~\cite{ATLAS:2023lsr,CMS:2024ony,ATLAS:2024erm,CMS:2024lrd} 
show that the results of determining precision SM parameters from the LHC data, such as 
the strong coupling constant, the effective leptonic weak mixing angle, 
and the $W$ boson mass, are highly sensitive to the choice of PDFs, 
with variations larger than the quoted PDF uncertainties. 
This suggests possible inconsistencies between independent PDF determinations 
and raises concerns about whether all sources of uncertainty are properly considered.
%
In particular it is not clear how these methods are able to robustly assess 
the dependence of the outcome of a fit on the choice of the prior (and further, the prior itself may not be well-defined or easy to describe in these approaches), 
and whether they may struggle when strong non-linear dependencies 
in the forward map are present~\cite{Costantini:2024wby}, thus motivating the exploration of 
a different approach, especially in view of simultaneous fits of PDFs and SMEFT Wilson coefficients~\cite{Costantini:2024xae,Kassabov:2023hbm,Iranipour:2022iak}.

Explicit Bayesian methods offer a fully articulated probabilistic treatment of inverse problems.
They have the advantage of providing quantitative estimates of all 
sources of uncertainties entering a PDF fit, particularly the methodological aspects. They also have the distinct benefit of making clear the choice of prior that is used in a PDF fit.
Interest in such approaches has recently increased in the literature, with several studies paving the way towards Bayesian PDF fits~\cite{Gbedo:2017eyp,Aggarwal:2022cki,Albert:2024zsh,Costantini:2024wby,Capel:2024qkm}. 
In particular, a promising approach is presented in Ref.~\cite{Candido:2024hjt}, in which a prior probability distribution 
for the PDF model is given by a Gaussian Process in the PDF functional space. 

On the other hand, Bayesian methods frequently invoke significant computational costs, 
especially when dealing with complex and high-dimensional PDF parametrisations, 
and it is not straightforward to apply them to a fully global PDF analysis, 
in which seven or eight flavours are extracted simultaneously from the 
${\cal O}(5000)$ data points involving both DIS and hadronic observables, 
where PDFs enter linearly and quadratically respectively.

In this work, we introduce a novel PDF parametrisation enabling a fully realistic Bayesian determination of the proton’s PDFs. 
Specifically, we parametrise PDFs as vectors in a linear space spanned 
by a small set of carefully chosen basis functions, which themselves are obtained by dimensional reduction of a neural network 
functional space through Proper Orthogonal Decomposition (POD). The dimensionality of this space can be dynamically adjusted, allowing for controlled 
flexibility in the parametrisation, i.e. increasing the number of basis functions increases the model’s adaptability.

The paper is organised as follows. In Sect.~\ref{sec:methodology}, we motivate and introduce the POD procedure allowing for the dimensional reduction of a 
PDF space to a linear functional space, where the basis vectors are organised in order of importance.
In Sect.~\ref{sec:basis}, we discuss the construction of a particularly versatile basis derived from the POD of a neural network functional space.
In Sect.~\ref{sec:model-selection}, we discuss the Bayesian workflow for model selection and model averaging, allowing us to carefully guard against 
underfitting and overfitting from either under-parametrised or over-parametrised PDF models, and describe the delivery of the PDF sets 
determined from the posterior distributions of the selected models. 
In Sect.~\ref{sec:results}, we present the results of a closure test fit to synthetic DIS data. We also perform a multi-closure test to validate the 
uncertainties of the resulting PDFs within the data region. This demonstrates that, by employing Bayesian model selection, we obtain a normalised bias 
compatible with unity within the bootstrap uncertainty.
Finally, in Sect.~\ref{sec:conclusion}, we summarise our findings and outline future directions, including the production of PDF fits based on real datasets. We also discuss the forthcoming public release of the code that was used to produce the results in this work, as a general-purpose, flexible, fast PDF-fitting platform, \texttt{colibri}~\cite{Costantini:2025agd}.

\section{Linear PDF models through proper orthogonal decomposition}
\label{sec:methodology}

Parton distributions at the initial scale are infinite-dimensional objects, with each flavour drawn from the space $C^1[0,1]$ 
of once-continuously differentiable functions on the unit interval, representing the fraction of a proton's momentum $x$ carried by 
a parton of a given flavour; as a result, they cannot be uniquely determined from a finite dataset. The natural solution to this problem
is to choose a parametric form for the PDFs, with a finite number of parameters, 
carving out a surface of possible PDFs in $C^1[0,1]^{N_{\rm flav}}$, where $N_{\rm flav}$ is the number of active 
flavours at the initial scale $Q_0$\footnote{Throughout this work, we will assume that $N_{\rm flav}=8$, hence we parametrise the gluon, up, anti-up, down, anti-down, strange, anti-strange as 
well as the charm PDF, which has both a perturbative component generated from the evolution of light quark/antiquarks and the gluon,  
and an intrinsic component; see Refs.~\cite{Brodsky:1980pb,Brodsky:2015fna,Ball:2022qks,Ball:2016neh,NNPDF:2023tyk} for a detailed discussion.}.


A choice of PDF parametrisation may have advantages and disadvantages. Three particularly desirable qualities of a PDF parametrisation are the following:
\begin{enumerate}[label = (\roman*)]
    \item The parametrisation should respect theoretical constraints, including small- and large-$x$ scaling behaviour, sum rules, and integrability.  
    \item The parametrisation should be sufficiently flexible to explore the space of candidate PDFs within the space $C^1[0,1]^{N_{\text{flav}}}$. 
    Furthermore, it should be possible to increase or decrease the model complexity easily, to allow for a data-driven determination of the appropriate 
    model complexity that describes the data.
    \item The parametrisation should enable efficient fitting of model parameters. This aspect is particularly critical for Bayesian methodologies, where 
    determining the posterior distribution requires exploring the parameter space using Markov Chain Monte Carlo (MCMC) methods, or other advanced Machine 
    Learning algorithms. The computational cost of these methods can become prohibitive when the parameter space is high-dimensional.  
\end{enumerate}

In order to perform a realistic PDF fit using a fully Bayesian methodology, we will focus on third quality, expedience of fitting 
-- something that has previously not been optimised for in the PDF literature -- without compromising on the first and second qualities either. 

\begin{figure}[t]
\centering
\begin{framed}
\begin{tikzpicture}[node distance=1cm and 1cm]

\node (start) [roundednode] {Identify a candidate space of PDFs, $\mathcal{H}$.};
\node (process) [roundednode, below=of start] {Perform POD to find optimal basis $\{\varphi_k(x)\}$ spanning a linear approximation of $\mathcal{H}$.};
\node (end) [roundednode, below=of process] {Perform a Bayesian fit to data using the parametrisation $f_{\vec{w}}(x) = \vec{w}^T\pmb{\varphi}(x)$.};

\draw [arrow] (start) -- (process);
\draw [arrow] (process) -- (end);

\end{tikzpicture}
\end{framed}
\caption{The workflow of the methodology we propose in this work. Sect.~\ref{sec:methodology} covers the second step, 
the proper orthogonal decomposition facilitating the reduction of a candidate PDF space to a linear functional space, 
spanned by an `optimal' choice of basis vectors in a sense to be defined in the text. Sect.~\ref{sec:basis} discusses 
the first step, namely the construction of the candidate space of PDFs. Sect.~\ref{sec:model-selection} discusses the third step, 
namely the Bayesian fit, its delivery, as well as model selection and Bayesian averaging.}
\label{fig:workflow}
\end{figure}
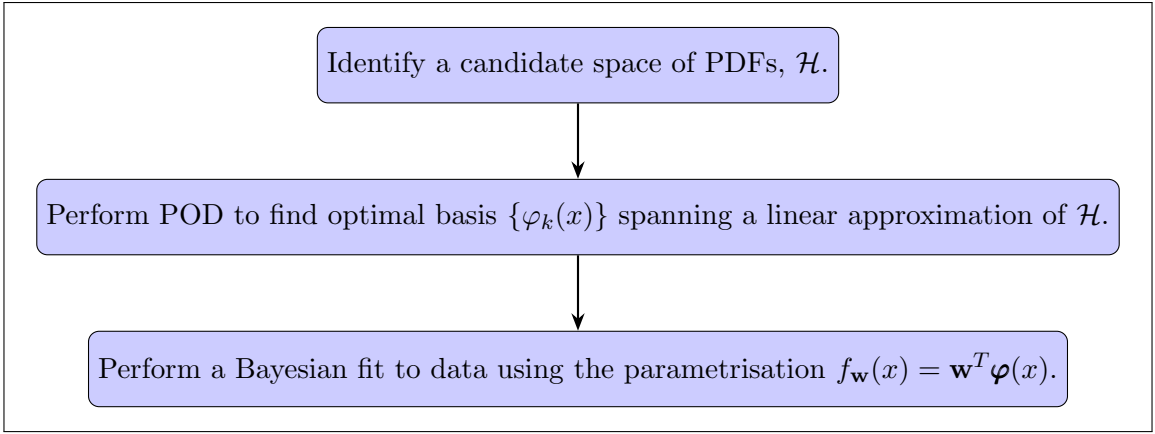
The workflow of our approach is summarised in Fig.~\ref{fig:workflow}. 
In this section, we introduce a new parametrisation, which is the `simplest possible' in terms of fitting: a \textit{linear model}, 
while the specific basis and details on the Bayesian fit will be discussed in the next sections.  
The linear model parametrises PDFs at the initial scale using the form  $f_{\vec{w}}(x) = \vec{w}^T \boldsymbol{\varphi}(x)$, where $\vec{w} = (1,w_1,...,w_N)$ 
is a finite collection of parameters, and $\boldsymbol{\varphi} = (\varphi_0(x),\varphi_1(x), ..., \varphi_N(x))$ is a collection of \textit{basis functions}. 
In Sect.~\ref{subsec:linear_pdf_model}, we describe the theory of how the basis functions can be obtained in an optimal fashion, given a candidate space of PDFs, 
such that the resulting linear functional form is sufficiently flexible to explore the relevant space of PDFs. Further, with this choice of basis functions, 
we go on to show that key theoretical constraints are satisfied by the resulting parametrisation.
Once we have established these constraints, in Sect.~\ref{subsec:pod} we shall move to the explicit construction of a particularly useful set of basis functions 
satisfying these conditions, given an initial candidate PDF space, through a method called \textit{proper orthogonal decomposition}.

\subsection{Constructing a linear PDF model}
\label{subsec:linear_pdf_model}
As described in the introduction to this section, the simplest possible PDF parametrisation that one could adopt is a \textit{linear model}:
\begin{align}
f_{\vec{w}}(x) &= \vec{w}^T \boldsymbol{\varphi}(x) = \varphi_0(x) + \sum_{k=1}^{N} w_k \varphi_k(x), 
\label{eq:linear_model}
\end{align}
for a finite collection of parameters $\vec{w} = (1,w_1,...,w_N)$ which we call \textit{weights}, and a set of $x$-dependent \textit{basis functions} $\boldsymbol{\varphi} = (\varphi_0(x), \varphi_1(x),...,\varphi_N(x))$ which span some vector subspace $S \subseteq C^1[0,1]^{N_\text{flav}} - \varphi_0$.\footnote{Here, $C^1[0,1]^{N_{\rm flav}} - \varphi_0 := \{\varphi - \varphi_0 : \varphi \in C^1[0,1]^{N_{\rm flav}}\}$.} Note that we have included a constant weight $1$ as the zeroth weight, to allow for some overall 
additive function $\varphi_0(x)$; this can often be beneficial in the fit of linear models (we shall see that it will be crucial in the implementation of \textit{sum rules} in the case of PDFs).

\paragraph{Theoretical constraints on the model.}
Before discussing the choice of the basis functions, it is important to address the consequences of choosing such a parametrisation for the theory constraints for the resulting PDF $f_{\vec{w}}(x)$. 
For this discussion, we shall write PDFs as column vectors in the $N_{\rm flav}$-dimensional PDF space,\footnote{Note that throughout this study, unless otherwise stated, we work in the evolution basis (described in more detail in Ref. \cite{NNPDF:2021njg}), and hence parametrise the PDF in terms of
\begin{align}
    \{\Sigma, g, V, V_3, V_8, T_3, T_8, T_{15}\} \;,
\end{align}
where $\Sigma$ is the singlet, $V$, $V_3$, $V_8$ are the non-singlet valence PDFs, 
and $T_3$, $T_8$, $T_{15}$ are the non-singlet triplet PDFs.} so that:
\begin{equation*}
f_{\vec{w}}(x) = \begin{pmatrix} f_{\vec{w}}^{\Sigma}(x) \\[1ex] f_{\vec{w}}^{g}(x) \\[1ex] \vdots \\[1ex] f_{\vec{w}}^{T_{15}}(x) \end{pmatrix}, \qquad \varphi_i(x) = \begin{pmatrix} \varphi_i^{\Sigma}(x) \\[1ex] \varphi_i^{g}(x) \\[1ex] \vdots \\[1ex] \varphi_i^{T_{15}}(x) \end{pmatrix}.
\end{equation*}
Then, the key theory constraints that arise on the PDF $f_{\vec{w}}(x)$ are the following:
\begin{itemize}
    \item \textbf{Valence sum rules.} The valence sum rules state that for each of the flavours $j = V, V_3, V_8$, the PDF must satisfy the following integral relation:
    \begin{equation*}
        \int\limits_{0}^{1} dx\  f_{\vec{w}}^j(x) = n_j,
    \end{equation*}
    where $n_V = n_{V_8}= 3, n_{V_3} = 1$, for all values of the parameter $\vec{w}$. 
    Inserting the linear parametrisation, we see that this relation implies:
    \begin{equation*}
        \int\limits_{0}^{1} dx\ \varphi^j_0(x) + \sum_{k=1}^{N} w_k \int\limits_{0}^{1} dx\ \varphi^j_k(x) = n_j
    \end{equation*}
    for all values of the parameter $\vec{w}$. Comparing coefficients, we see that 
    %
    the basis element $\varphi_0(x)$ satisfies the usual valence sum rules for PDFs, but the remaining basis elements $\varphi_k(x)$ with $k\ge 1$ satisfy `homogeneous' 
    versions of the valence sum rules. That is, each of the basis elements $\varphi_k(x)$ must satisfy sum rules where $n_j$ is replaced by zero everywhere.

    \item \textbf{Momentum sum rule.} The momentum sum rule states that:
    \begin{equation}
        \int\limits_{0}^{1} dx\ x \left( f_{\vec{w}}^{\Sigma}(x) + f_{\vec{w}}^{g}(x)\right) = 1,
    \end{equation}
    for all values of the parameter $\vec{w}$. 
    In exactly the same way, this implies that the basis elements must satisfy:
    \begin{equation}
       \int\limits_{0}^{1} dx\ x \left( \varphi^{\Sigma}_0(x) + \varphi^{g}_0(x) \right) = 1, \qquad \int\limits_{0}^{1} dx\ x \left( \varphi_k^{\Sigma}(x) + \varphi_k^{g}(x) \right) = 0,
    \end{equation}
    so that the basis element $\varphi_0(x)$ satisfies the usual momentum sum rule for PDFs, but the remaining basis elements $\varphi_k(x)$ satisfy a `homogeneous' version of the momentum sum rule.

    \item \textbf{Integrability.} The third key theory constraint we would like to impose on the model is \textit{integrability}, which ensures that certain flavour combinations are integrable. 
    In the evolution basis, this requires that:
    \begin{align*}
        & \lim_{x \rightarrow 0} x^2 f^j_{\vec{w}}(x) = 0,\qquad j=g,\Sigma \\
       & \lim_{x \rightarrow 0} x \,f^j_{\vec{w}}(x) = 0, \qquad\ j=V,V_3,V_8 \; .
    \end{align*}
    Furthermore, standard Regge theory arguments suggest that the first moments of the non-singlet triplet combinations $T_3$ and $T_8$ are also finite, so for instance the Gottfried 
    sum rule (which is proportional to the first moment of $T_3$) is finite. This implies that also for these two combinations we have:
    \begin{align*}
       & \lim_{x \rightarrow 0} x f^j_{\vec{w}}(x) = 0,\qquad \ j=T_3,T_8.
    \end{align*}
    Since these are all linear, homogeneous conditions, we see once again that if all the basis elements $\varphi_0(x)$, $\varphi_k(x)$ satisfy the integrability conditions, then $f_{\vec{w}}(x)$ will satisfy the integrability conditions.
\end{itemize}
We shall show in the next section that using the method of \textit{proper orthogonal decomposition} produces a basis $\varphi_k(x)$ with these properties.

\subsection{Proper orthogonal decomposition}
\label{subsec:pod}
There are many ways to choose an $N$-dimensional basis for representing PDFs once they have been discretised on a grid, i.e.\ as vectors in $\mathbb{R}^{N_{\rm flav}N_{\rm grid}}$.
For example, one may expand each flavour in orthogonal (e.g., Chebyshev~\cite{Martin:2012da}) polynomials and then evaluate the result on the grid, or use B\`{e}zier parametrizations~\cite{Kotz:2025une}.
However, such generic choices are not necessarily the most efficient for representing realistic PDFs.
For fitting it is therefore useful to \quotes{design} a basis that captures the typical behaviour of the candidate ensemble.
Proper orthogonal decomposition (POD) provides a simple, optimal way to do this.

We work directly in the discretised setting relevant for the numerical PDF fit.
We start from a large ensemble of $M$ candidate PDFs, drawn from an initial sampling that spans a broad region of the full PDF functional space. The ensemble size M is assumed to be much larger than $N$, ensuring a sufficiently rich representation of functional variability.
More details on how $M$ is chosen are discussed in Section~\ref{sec:basis}.
The candidate PDF functions drawn from the initial ensemble have been evaluated on an $x$-grid of size $N_{\rm grid}$ and stacked over flavours.
Each ensemble member is therefore represented as a $\mathbb{R}^{N_{\rm flav}N_{\rm grid}}$-dimensional vector in Euclidean space
\begin{equation}
    \vec{g}_m := \begin{pmatrix} g_m^{\Sigma}(x_1) \\[1ex] \vdots \\ g_m^{\Sigma}(x_{N_{\rm grid}}) \\[1ex] g_m^{g}(x_1) \\ \vdots \end{pmatrix}
    \in \mathbb{R}^{N_{\rm flav}N_{\rm grid}} .
\end{equation}

\paragraph{Construction of the POD modes.}
We first centre the ensemble on its mean\footnote{The choice of centring is not unique. In principle, the POD construction yields a valid PDF parametrisation for any choice of reference function, provided it corresponds to a physically admissible PDF candidate.},
\begin{equation}
    \pmb{\varphi}_0 := \frac{1}{M}\sum_{m=1}^M \vec{g}_m, 
    \qquad
    \tilde{\vec{g}}_m := \vec{g}_m - \pmb{\varphi}_0 ,
\end{equation}
and construct the $N_{\rm flav}N_{\rm grid} \times N_{\rm flav}N_{\rm grid}$ autocorrelation matrix
\begin{equation}
    A = \sum_{m=1}^M \tilde{\vec{g}}_m \tilde{\vec{g}}_m^T .
\end{equation}
Since $A$ is symmetric and positive semi-definite, it admits an orthonormal eigenbasis $\{\vec{e}_k\}$ with eigenvalues
$\lambda_1 \ge \lambda_2 \ge \cdots \ge 0$.
The POD modes are precisely these eigenvectors (up to an overall normalisation convention),
\begin{equation}
    A \vec{e}_k = \lambda_k \vec{e}_k , \qquad \pmb{\varphi}_k \propto \vec{e}_k,
\end{equation}
where $k=1,\ldots,N_{\rm flav}N_{\rm grid}$. Truncating after $N$ modes retains the directions with the largest eigenvalues and therefore captures the dominant variance of the ensemble.
Equivalently, the subspace spanned by the first $N$ modes yields the best rank-$N$ approximation among all $N$-dimensional subspaces \cite{Holmes_Lumley_Berkooz_Rowley_2012}.
This makes POD efficient: often only a few modes are needed in practice \cite{Carrazza:2015aoa, Gao:2013bia, Courtoy:2022ocu}.

\paragraph{Discretised POD parametrisation.}
We use these POD modes directly as the PDF parametrisation. The discretised analogue of Eq.~\eqref{eq:linear_model} is
\begin{equation}
    \vec{f}_{\vec{w}} := \pmb{\varphi}_0 + \sum_{k=1}^{N} w_k\,\pmb{\varphi}_k
\label{eq:linear_model_discrete}
\end{equation}
where $\vec{f}_{\vec{w}}\in\mathbb{R}^{N_{\rm flav}N_{\rm grid}}$ is the stacked PDF vector evaluated on the grid, $\pmb{\varphi}_0$ the mean of the ensemble, 
$\pmb{\varphi}_k$ the centred POD modes and $\vec{w}=(1,w_1,\dots,w_N)$ are the free parameters of the PDF parametrisation.

Computationally, this finite-dimensional POD is equivalent to performing a singular value decomposition (SVD) of the centred data matrix; 
this provides a standard and numerically stable implementation (see Sect.~3.4.2 of Ref.~\cite{Holmes_Lumley_Berkooz_Rowley_2012}).  
By construction, any linear and homogeneous constraints satisfied by the ensemble are preserved by the resulting basis (see App.~\ref{app-infinite_dimensional_pod}).

In summary, POD yields a low-dimensional linear basis that is optimal for approximating a given ensemble of PDFs while respecting all relevant theory constraints.  
In the next section we specify the starting ensemble of size $M$ on which the POD is performed.

\section{Building a basis for the PDF functional space}
\label{sec:basis}

In Sect.~\ref{sec:methodology}, we described the use of POD for the reduction of a candidate space of PDFs to a linear approximation of the space. 
However, we remained deliberately vague about the exact choice of candidate PDF space; the selection of this space is a 
crucial step, as it determines the functional space that the model can ultimately span. 
Since our goal is to capture a broad region of the complete PDF functional space, 
this choice plays a key role in shaping the model’s capabilities. A limited or overly 
restrictive candidate PDF space would constrain the flexibility of the resulting POD model, 
potentially reducing its effectiveness in describing the data.

In the following subsections, 
we choose to work with an initial candidate PDF space parametrised by a deep neural network (NN) architecture.
From this space, we draw a large number of samples from random initialisation of the NN, on which we perform POD to construct a linear functional 
space which approximates the candidate PDF space.
In a numerical study, we proceed to prove that the resulting POD approximation to the NN space provides: 
(i) a complete approximation, in the sense that any PDF replica produced using a random initialisation of the NN can be 
well-approximated by the resulting linear model; (ii) a general parametrisation, in the sense that PDF replicas drawn from other 
parametrisations (for example the latest parametrisation used by the NNPDF, MSHT and CT collaborations) can also be well-approximated by the resulting linear model.

The structure of this section is as follows. We begin in Sect.~\ref{subsec:nn} with a discussion of the exact NN space that we will use as our initial candidate PDF space. 
In Sect.~\ref{subsec:completeness} we proceed to discuss the POD reduction of this NN space, and prove that it satisfies the completeness and flexibility conditions we described above.

\subsection{The Neural Network POD Basis}
\label{subsec:nn}
The use of a neural network PDF space as our initial candidate PDF space was not our first choice; instead, 
we began by using a large sample of PDF replicas drawn from a variety of PDF sets from the LHAPDF repository~\cite{Buckley:2014ana}. 
Na\"{i}vely, one might expect that this might be the best choice as a candidate PDF space, as it would allow for the reduction 
of `known PDF space', as explored by all existing PDF collaborations to date.  The main problem arising from such an approach is that 
this space of PDFs has already been trained on data, and as such induces a strong bias in the resulting POD space, hampering the generalisation capabilities of the basis. 
As a result, we moved to study a space of PDFs which had not already seen data.

An approach which avoids the data-induced bias is to instead start from a candidate PDF space parametrised by a deep enough NN, which is well known to be a 
universal approximant in the limit of a sufficiently deep network. The use of NNs to fit PDFs 
was pioneered by the NNPDF collaboration a few decades ago~\cite{Ball:2008by,Ball:2010de}. Here we choose to use a deep NN with the same architecture as the one selected 
in the most recent NNPDF collaboration global analysis, NNPDF4.0~\cite{NNPDF:2021njg,Carrazza:2019mzf},  
by means of a hyper-optimisation procedure~\cite{Cruz-Martinez:2024wiu}. We then draw a sample from this neural network space, so that in the 
language of Sect.~\ref{sec:methodology}, each flavour of our initial sample from the space has the functional form:
\begin{align} 
x g_m^j(x, Q_0; \boldsymbol{\theta}) = A^j_m x^{1-\alpha^j_m}(1-x)^{\beta^j_m} \text{NN}_m(x; \boldsymbol{\theta}),
\label{eq:neural_net_param}
\end{align}
where $m=1,...,M$ runs over the elements of the initial sample, $j=1,...,N_{\rm flav}$ runs over the flavour indices, 
and $\text{NN}_m(x; \boldsymbol{\theta})$ represents the $m$-th output of a
fully connected feed-forward neural network with an input layer of 2 neurons, followed by two hidden layers with 25 and 20 neurons, 
respectively and with $N_{\rm flav}$ outputs. Each neuron, except the output layer which is linear, has $\tanh$ 
activation function and the network weights are initialized using the Glorot normal distribution \cite{pmlr-v9-glorot10a}. 

\paragraph{Generation and validation of the sample.} The sample generation proceeds by randomly initializing the NN 
weights according to the Glorot distribution, and uniformly sampling the exponents $\alpha_m^j$ and $\beta_m^j$ within 
the ranges specified in \cite{NNPDF:2021njg}. Note that this is just one of the many possible choices of the exponent range, 
which we pick for reference. This is actually a delicate point, as the choice of pre-processing can significantly influence the resulting functional space. It is therefore important to assess this choice carefully, ideally using data-driven criteria. A thorough investigation of this aspect is left for future work.
The normalisation constants $A_m^j$ are then determined by enforcing the valence and momentum sum rules, 
as discussed in Sect. 3.1.2 of \cite{NNPDF:2021njg}. Each PDF sample, $g^{j}_m$, represents a possible 
realisation of the pre-processed NN at initialisation, and is stored as a vector in a $N_{\rm grid} \times N_{\rm flav}$ dimensional space.

The resulting sample may include replicas with very large arc lengths when the sum rule normalisation factor is close to zero. 
Since such replicas are not physically reasonable, we apply an outlier removal procedure to ensure a well-behaved basis. 
Specifically, we compute the arc length of each replica and remove extreme outliers using an interquartile range (IQR) filter. 
This step helps eliminate replicas that introduce spurious fluctuations.

\begin{figure}[t!]
    \centering
    \includegraphics[width=0.75\linewidth]{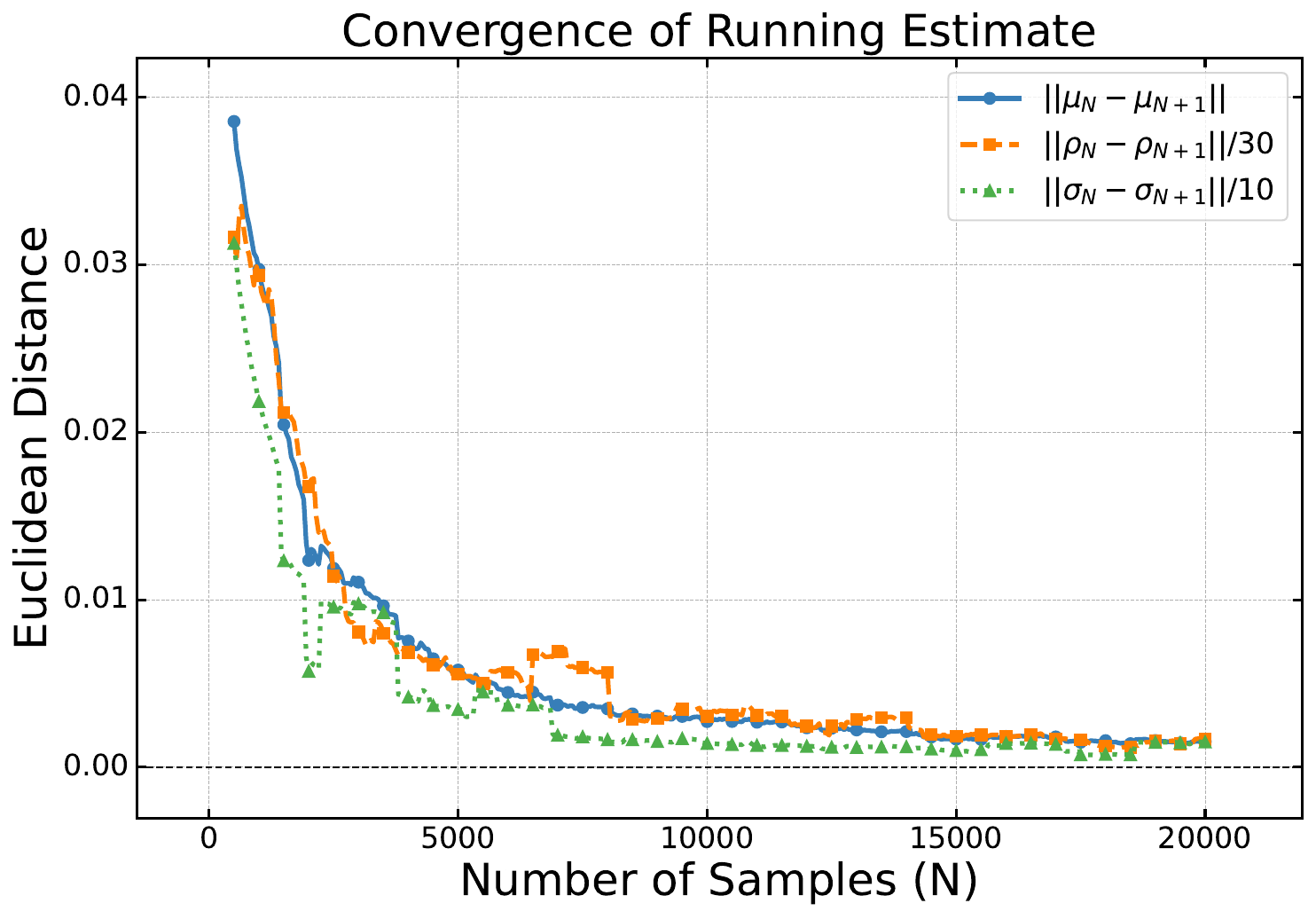}
    \caption{Convergence of the mean (blue solid line), variance (green dotted line) and correlation (orange dashed line) 
    of the NN probability density distribution, estimated in terms of the rolling averaged Euclidean distance between estimates based on $M$ and $M+1$ samples.}
    \label{fig:nn_distribution_convergence}
\end{figure}

%
Finally, before applying POD to the PDF sample obtained from random realisations of our NN, 
we must ensure that we have a sufficient number of samples to accurately estimate 
the density of the neural network probability density distribution, which is determined by the NN architecture of Eq.~\eqref{eq:neural_net_param} 
and by the distribution of its weights and biases. 
Fig.~\ref{fig:nn_distribution_convergence} illustrates the convergence of the neural network distribution's mean, variance, 
and correlation. Specifically, it shows how the rolling averaged Euclidean distance between estimates based on $M$ and $M+1$ samples 
varies with $M$. For all estimates, these distances decrease rapidly and flatten to values close to zero well before $1\cdot 10^4$ samples. 
Based on this observation, we choose to construct the basis using $M=2\cdot 10^4$ initial samples.

\subsection{Assessing the completeness of basis and the flexibility of the model}
\label{subsec:completeness}

Starting from the large initial sample of  $\{g_1(x), ..., g_M(x)\}$ with $M=2\cdot 10^4$, 
we use POD, as described in Sect.~\ref{subsec:pod}, to reduce to a basis $\{\varphi_1(x), ..., \varphi_N(x)\}$. 
In this section, we validate our methodology by performing two tests that assess whether the starting parametrisation and its reduced POD representations  
are sufficiently flexible to capture the full range of PDF behaviours that one might encounter in a realistic PDF fit. 
We test two different aspects of our basis: its completeness and its generalisation capability. 
Completeness is defined as the ability of the POD parametrisation to reproduce any random realisation of the original parametrisation 
that was used to produce the POD basis; this essentially tests how closely the linear POD parametrisation approximates the original, generally non-linear, parametrisation. 
The generalisation capability of the basis is defined as its ability to reproduce a PDF representation that does not come from the original parametrisation, 
but still satisfies all the theoretical constraints that a PDF replica should have. If the original space was sufficiently broad as to be able to 
describe many possible PDFs, then the POD reduction should inherit this quality too.

In order to test both features, we consider the mean squared error between PDFs, evaluated at specific $x$-grid points as
\begin{equation}
\text{MSE} = \frac{1}{N_{\rm targ}} \sum_{i=1}^{N_{\rm targ}}\left\| \mathbf{f}^{(i)}_T - \mathbf{f}_{\vec{w}^{(i)}} \right\|_2^2 
  = \frac{1}{N_{\rm targ}} \sum_{i=1}^{N_{\rm targ}} \left\| \mathbf{f}^{(i)}_T - \left( \pmb{\varphi}_0 + \sum_{n=1}^N 
  w_n^{(i)} \pmb{\varphi}_k \right) \right\|_2^2 \, ,
\label{eq:euclid_dist_between_pdfs}
\end{equation}
where $\mathbf{f}^{(i)}_T$ is the $i$-th member of a target ensemble of PDFs that we wish to reproduce, 
and $N_{\rm targ}$ is the total number of PDFs in the target ensemble. The vector $\vec{f}_{\vec{w}^{(i)}}$ is a function of 
the weights $\vec{w}^{(i)}$, which will be different for different target replicas, and are chosen to minimise the mean-squared 
error MSE. The second equality displays the definition of $\vec{f}_{\vec{w}}$ from the previous section. The PDF vectors $\vec{f}_T^{(i)}$ 
and $\vec{f}_{\vec{w}}^{(i)}$ are evaluated on the points of the standard LHAPDF \cite{Buckley:2014ana} grid that are $\geq 10^{-5}$; these PDFs are given at the parametrisation scale $Q_0 = 1.65\ \mathrm{GeV}$, although in principle the same exercise can be carried out at any choice of scale $Q$.
Further, they represent \textit{stacked} vectors of flavours, so that for each $(i)$, $\vec{f}^{(i)}_T$ represents each of the $N_{\rm flav}$ flavours of a particular 
target evaluated on the LHAPDF grid, stacked as a vector.

\begin{figure}[t!]
    \centering
    \includegraphics[width=0.49\linewidth]{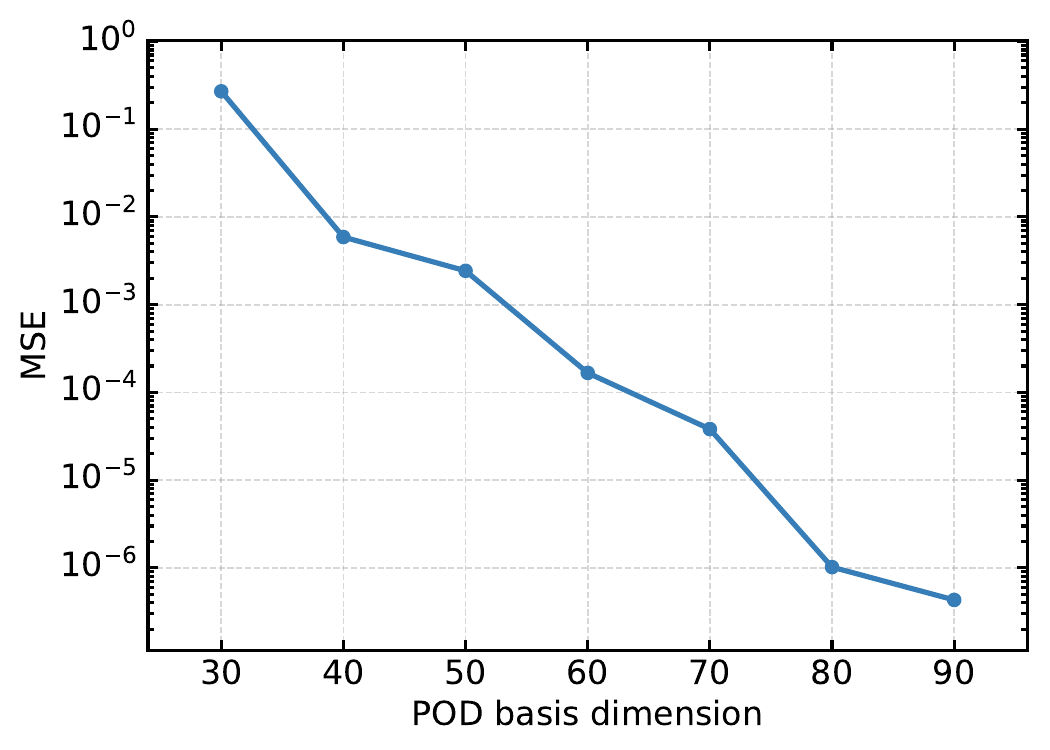}
    \includegraphics[width=0.49\linewidth]{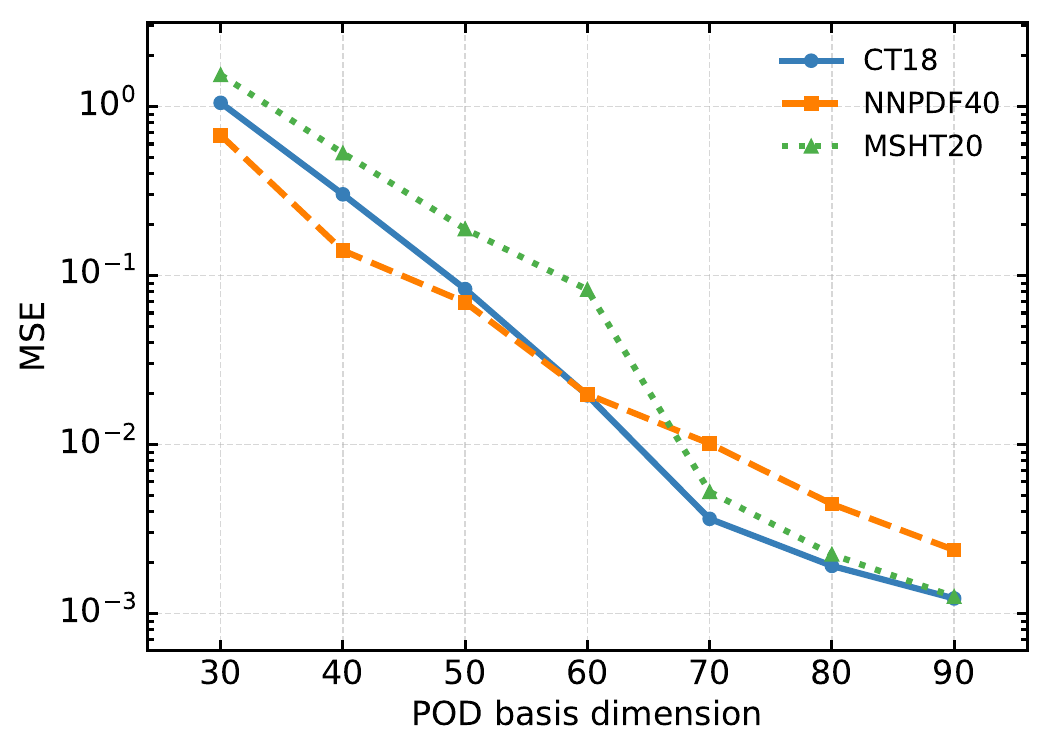}
    \caption{Left panel: we test the completeness by plotting the MSE between fixed-basis-dimension POD model and 
    100 randomly drawn realisation of a NN, Eq. \eqref{eq:neural_net_param}, as a function of the POD basis dimension $N$. 
    Right panel: we test the generalisation by plotting the MSE  between the POD model and three alternative PDF parametrisation  
    from the CT18 NNLO PDF set~\cite{Hou:2019efy} (blue solid line), the MSHT20 NNLO PDF set~\cite{Bailey:2020ooq} (green dotted line), 
    and the NNPDF4.0 NNLO PDF set~\cite{NNPDF:2021njg} (orange dashed line).}
    \label{fig:mse_dist}
\end{figure}

\begin{figure}[t!]
\centering
\includegraphics[width=0.474\textwidth, page=1]{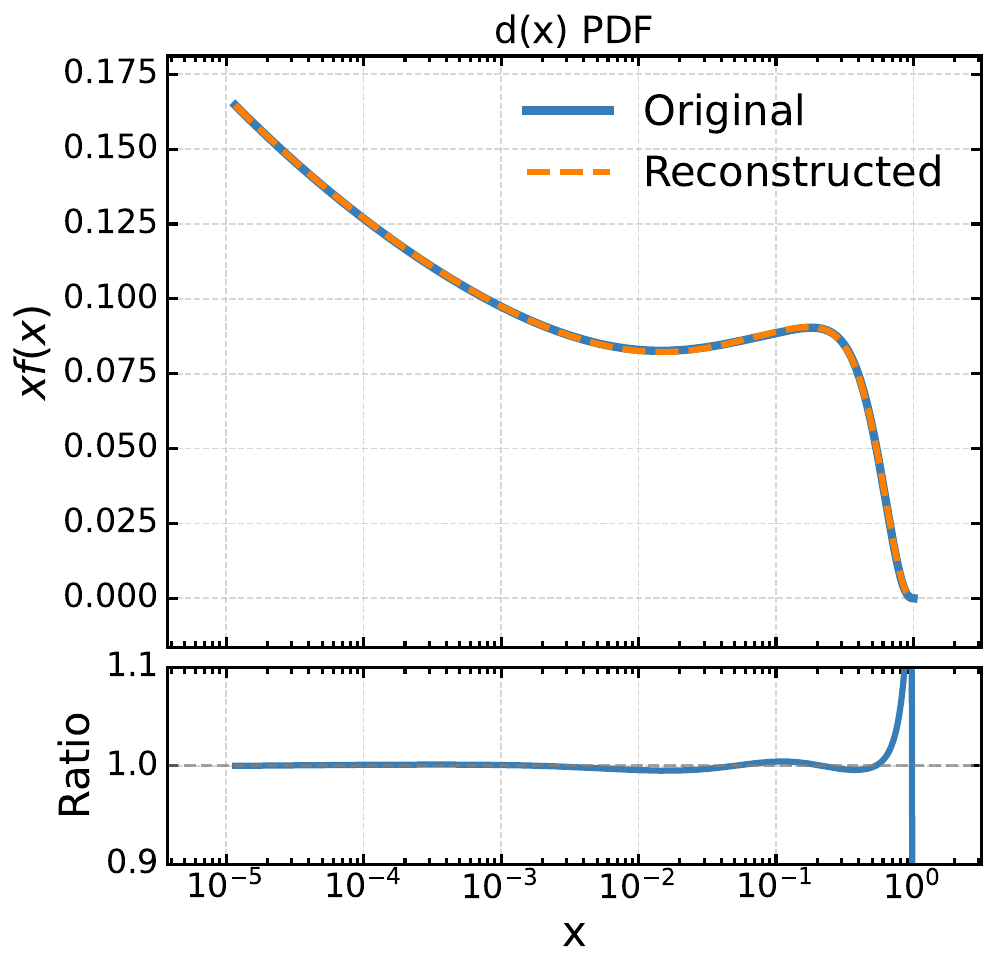}
\includegraphics[width=0.45\textwidth, page=8]{Figures/plots_PDF_50w}
\caption{Reconstruction of random realisation of a NN, Eq.~\eqref{eq:neural_net_param}, 
using a POD parametrisation with $N=50$ basis members. \label{fig:pdf_target_plots}}
\end{figure}

\paragraph{Completeness.} To assess the completeness of the basis, we draw a random sample of replicas from the same parametric form used to generate the POD basis. 
Importantly, we ensure that the random seeds used differ from those employed in constructing the basis, to avoid any overlap or bias.
The left hand side of Fig.~\ref{fig:mse_dist} shows the mean squared distance between a fixed-basis-dimension POD model and 100 random replicas generated from Eq.~\eqref{eq:neural_net_param}.
 The total distance between the target and reconstructed replicas has an approximately exponential decay with increasing dimensionality of the basis. 
 This reflects the fact that the POD basis is ordered such that the most important modes are used first, followed by successively less important modes. 
 To gain intuition on the meaning of the squared distance shown on the $y$-axis of the plot, in Fig.~\ref{fig:pdf_target_plots}, we illustrate 
 how a POD model with 50 components, corresponding to an MSE value of $\sim 10^{-3}$, can accurately reconstruct a randomly sampled replica.

\paragraph{Generalisation.} In order to test the generalisation ability of the POD basis, we reproduce PDFs from a 
wide ensemble of PDF sets, collected in Table~\ref{tab:pdf_target}. The sets used are taken from three widely-used NNLO PDF 
sets CT18~\cite{Hou:2019efy}, NNPDF4.0~\cite{NNPDF:2021njg}, and MSHT20~\cite{Bailey:2020ooq}. 
\begin{table}[tb]
    \centering
    \caption{PDF sets and number of replicas that constitute the test target for the generalisation benchmark of the POD model.}
    \begin{tabular}{lc}
    \toprule
    PDF Set & Number of Members ($N_{\rm targ})$ \\
    \midrule
    CT18NNLO & 58 \\
    NNPDF40\_nnlo\_as\_01180 & 100 \\
    MSHT20nnlo\_as118 & 64 \\
    \bottomrule
    \end{tabular}
    \label{tab:pdf_target}
\end{table}
On the right-hand side of Fig.~\ref{fig:mse_dist} we show the mean squared distance computed on the three PDF 
sets of Table \ref{tab:pdf_target} as a function of the number of POD basis elements. All three sets can be reproduced well. 
Interestingly, the set NNPDF40, determined by fitting the same parameterisation as 
Eq.~\eqref{eq:neural_net_param}, has a similar MSE as CT18 and MSHT20. This is likely due to the fact that the distributions 
of the NN weights trained on the data were shifted from the initial Glorot distribution.
\begin{figure}[tb]
\centering
\includegraphics[width=0.45\textwidth, page=1]{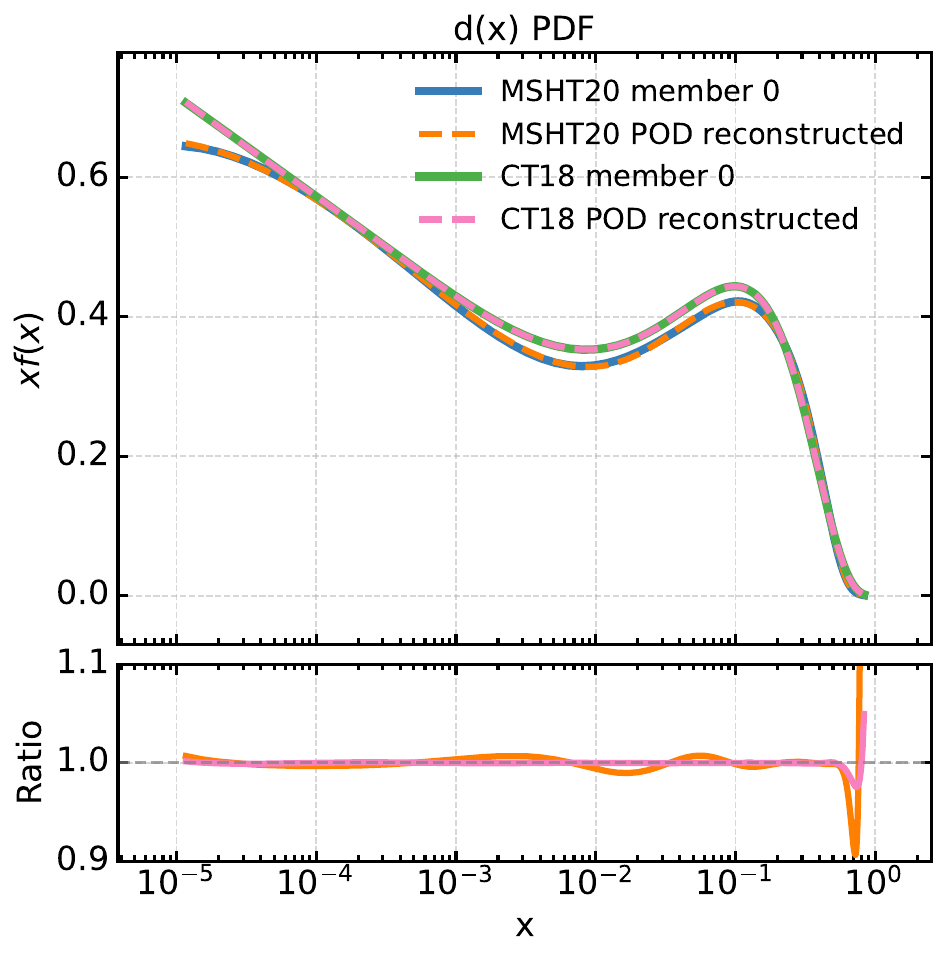}
\includegraphics[width=0.45\textwidth, page=8]{Figures/plots_PDF_msht_80w}
\caption{Reconstruction of random replica from the MSHT and CT18 PDF set using a POD parametrisation with N=80 basis members.\label{fig:pdf_target_plots_2}}
\end{figure}
In Fig.~\ref{fig:pdf_target_plots_2} we show the reconstruction of the central member of the MSHT20 and CT18 NNLO PDF sets, 
by using a basis dimension of 80 corresponding to a mean squared distance of approximately $10^{-3}$. As it can be gathered from the plot, 
the reconstruction is within 1\% of the target, with the exception of the large-$x$ region where the PDFs are very close to zero. 

\section{Fitting methodology and model selection}
\label{sec:model-selection}
%

After having presented the initial PDF functional space and its POD reduction, we can now turn to the problem of inferring the PDF parameters 
from the available data. 
%
In Sect.~\ref{subsec:likelihood}, we state the Bayesian posterior distribution, including contributions from the likelihood 
and from a prior requirement that PDFs satisfy certain additional theoretical constraints.
In Sect.~\ref{subsec:fitting_method}, we proceed to describe the implementation details of the numerical sampling algorithm in this work. 
In particular, we discuss how the fit can be accelerated through the use of an updating strategy, based on performing a simpler, analytic fit, 
before proceeding with the numerical fit. In Sect.~\ref{subsec:model_averaging}, we describe a Bayesian method of model selection 
and averaging, allowing us to assess how many parameters from our POD model are appropriate, and allowing us to include error from our 
uncertainty over the correct choice of PDF model. Finally, in Sect.~\ref{subsec:deliverable}, we discuss how the resulting PDF 
set is delivered once the posterior has been determined. 

\subsection{Likelihood function and theoretical constraints}
\label{subsec:likelihood}

In the collinear factorisation framework, the theoretical prediction for an observable is given by the convolution of the PDFs, 
$f$, with perturbative partonic cross sections $\hat\sigma$. In practice, this convolution is replaced by the multiplication by 
a pre-computed interpolation grid, so that the computation of theoretical observables is fast and efficient during the fit, 
see Ref.~\cite{Candido:2022tld} for a recent summary, describing also the code used in producing the interpolation tables, in our case the Fast Kernel (FK) tables used in this work. 
We denote this mapping from PDFs to data space as $\mathbf{T}[f]$, which is a vector of $N_{\text{data}}$ points, to be compared 
to the $N_{\text{data}}$ available data.

The experimental central data $\mathbf{D}$, a vector of $N_{\text{data}}$ points, are assumed to follow a multivariate normal distribution:
\begin{align}
\mathbf{D} \sim \mathcal{N}(\mathbf{T}[f^*], C),
\end{align}
where $C$ is the experimental covariance matrix, typically modified following the $t_0$ prescription when multiplicative uncertainties are present to avoid the d'Agostini bias~\cite{Ball:2009qv}.\footnote{Further, the covariance matrix may be modified to also include nuclear uncertainties or missing higher-order uncertainties, as discussed in Refs.~\cite{Ball:2018twp,NNPDF:2019vjt,NNPDF:2019ubu,NNPDF:2024dpb}. Some datasets used in this work automatically include nuclear uncertainties, for example the deuteron and NMC datasets.} 
The vector $\mathbf{T}[f^*]$ represents the theory prediction evaluated on the true PDF $f^*$. In real-life situation we do not have access to the 'true' underlying 
law of Nature, while in the the closure-test setting which follows in Sect.~\ref{sec:results}, we choose $f^*$ ourselves, and aim to recover it.

Given this setup, the likelihood of the experimental data $\vec{D}$ given that the correct PDF model is $f_{\vec{w}}$ with a fixed number of weights (an assumption we shall return to), 
is given by:
\begin{align}
p(\mathbf{D} | \mathbf{w}) = \frac{1}{(2\pi)^{N/2} |C|^{1/2}} 
\exp \left( -\frac{1}{2} \chi^2_{\rm data}(\mathbf{D}, \mathbf{w}) \right),
\end{align}
where
\begin{align}
\chi^2_{\rm data}(\mathbf{D}, \mathbf{w}) = (\mathbf{D} - \mathbf{T}[f_{\mathbf{w}}])^T C^{-1} (\mathbf{D} - \mathbf{T}[f_{\mathbf{w}}]),
\end{align}
and $|C|$ denotes the determinant of the covariance matrix. 

In addition to fitting the data, theoretical constraints must be imposed to ensure physically meaningful PDFs. In the Bayesian framework, 
these can be considered to be prior knowledge, which multiplies the likelihood described above. Assuming a normal form of the prior, 
this is mathematically equivalent to instead adding some \textit{penalty} terms to the $\chi^2$ featuring in the exponent of the likelihood, 
which is the view we shall take below.

As discussed in Sect.~\ref{sec:methodology}, the form of the POD parametrisation should in principle ensure some of the required theoretical constraints 
are automatically satisfied; in particular, it should be the case that both sum rules and integrability are enforced for our PDFs. On the other hand, 
another theoretical constraint - that physical cross sections produced from our PDFs are positive (and further, that the PDFs themselves are positive for individual quark and gluon flavours~\cite{Candido:2020yat, Candido:2023ujx}) - is \textit{not} something that our parametrisation should guarantee. In practice, however, we find it useful to 
impose both positivity of physical cross sections and PDFs, \textit{and} integrability as theoretical constraints. In more detail:

\paragraph{Positivity of physical cross sections and PDFs.} We enforce positivity both on physical cross sections and on 
PDFs themselves. For the former, we imposed positivity of some observables (in particular the flavour specific $F_2^u$, $F_2^d$, $F_2^s$, $F_2^c$ structure functions and   
the longitudinal structure function $F_L$) by adding positivity pseudo-observables and imposing that 
they are positive by means of a Lagrange-multiplier penalty term, exactly as it is done in Ref.~\cite{NNPDF:2021njg}. 
For the latter, we follow Refs.~\cite{Candido:2023ujx,Candido:2020yat}, where it is shown that PDFs for individual 
quark flavours and the gluon in the $\overline{MS}$ factorization scheme are non-negative in the medium-large $x$ region. 
To this purpose we add a Lagrange-multiplier penalty term to the $\chi^2$ to impose the positivity of the flavours $j\in\{u, \bar{u}, d, \bar{d}, s, \bar{s}, g\}$
\begin{align}
\chi^2_{\rm pos}(\mathbf{w}) = \Lambda_{\rm pos} \sum_j\sum_{\beta=1}^{N_{\rm grid}}
\operatorname{Elu}_{\alpha}\bigl(-\tilde{f}^j_{\mathbf{w}}(x_\beta,Q^{2})\bigr),
\end{align}
at $Q^2 = 5\,\mathrm{GeV}^2$, and $x_\beta$ are points 
on a $N_{\rm grid}$-dimensional grid in the range $0.1 \leq x \leq 0.9$. The function $\operatorname{Elu}_\alpha$ is given by
\begin{equation}
    \operatorname{Elu}_{\alpha} (t)= 
    \begin{cases}
        t & \text{if } t > 0 \\
        \alpha (e^t - 1) & \text{if } t < 0\; .
    \end{cases}
\end{equation}

\paragraph{Integrability.} Despite our expectation that our PDF parametrisation would automatically ensure integrability, we have found that the oscillatory behaviour
of our basis functions can induce an unexpected numerical instability in
evolution codes such as EKO~\cite{Candido:2022tld} and APFEL~\cite{Bertone:2016lga}. Such numerical
instability, in particular, might lead to a deterioration of the sum rules as a function of the evolution scale, which could become significant in a fit of real data. 
To mitigate this, we introduce the following additional penalty term on the small-$x$ behaviour of $T_3$ and $T_8$, preventing any numerical error from propagating 
through the evolution code:
\begin{align}
\chi^2_{\rm integ}(\mathbf{w}) = \Lambda_{\rm int} \sum_{j \in \{T_3, T_8\}} \sum_{\beta=1}^{N_{\rm grid}}
\bigl[x_\beta f^j_{\mathbf{w}}(x_\beta, Q_0^2)\bigr]^2,
\end{align}
where $Q_0$ is the parametrisation scale and $x_\beta$ covers the small-$x$ region of the grid. 
In the future, we hope that numerical evolution issues such as these can be reduced, removing the requirement to impose integrability as a penalty term. 


With the inclusion of these penalty terms, the complete likelihood considered in this study is then:
\begin{align}
p(\mathbf{D} | \mathbf{w}) \propto  
\exp \left( -\frac{1}{2} \bigl[\chi^2_{\rm data}(\mathbf{D}, \mathbf{w}) + \chi^2_{\rm pos}(\mathbf{w}) + \chi^2_{\rm integ}(\mathbf{w})\bigr] \right),
\label{eq:general_likelihood}
\end{align}
where $\vec{w}$ denotes the PDF parameters. This final form ensures that physical and theoretical constraints are consistently incorporated during inference.

\subsection{Fitting methodology}
\label{subsec:fitting_method}
As discussed above, the full likelihood used in the fit, Eq.~\eqref{eq:general_likelihood}, combines the agreement with data and the enforcement of theoretical constraints. The Bayesian posterior distribution for the PDF parameters $\vec{w}$ is then given by:
\begin{align}
\label{eq:posterior}
p(\mathbf{w} | \mathbf{D}) = \frac{p(\mathbf{D} | \mathbf{w}) p(\mathbf{w})}{\mathcal{Z}},
\end{align}
where $p(\vec{w})$ is our prior distribution on the weights, and $\mathcal{Z}$ is the evidence (or marginal likelihood) obtained by integrating 
the likelihood over the prior on $\mathbf{w}$. In general, the likelihood is non-Gaussian, and hence the Bayesian posterior is non-Gaussian; as a result, 
in the most general case, the posterior and the evidence are computed using the nested sampling algorithm as implemented in the \texttt{ultranest} 
package~\cite{2019PASP..131j8005B, 2021JOSS....6.3001B}. However, to improve computational efficiency, we adopt a Bayesian updating strategy 
that takes advantage of the fact that part of the parametric dependence on the $\mathbf{w}$ parameters is linear.

\paragraph{Bayesian updating strategy.} For computational efficiency, we exploit the fact that the dataset $\mathbf{D}$ can often be split 
into two uncorrelated subsets: $\mathbf{D}_1$ (e.g.\ the linear DIS data) and $\mathbf{D}_2$ (e.g.\ hadronic data or ratio data). This leads to a factorised likelihood:
\begin{align}
p(\mathbf{D} | \mathbf{w}) = p(\mathbf{D}_1 | \mathbf{w})\, p(\mathbf{D}_2 | \mathbf{w}),
\end{align}
where the covariance matrix $C = C_1 \oplus C_2$ is block-diagonal. 

When $\mathbf{D}_1$ corresponds to data that are linear in $\mathbf{w}$ (as in the case of POD parametrisations), the likelihood $p(\mathbf{D}_1 | \mathbf{w})$ is Gaussian:
\begin{align}
p(\mathbf{D}_1 | \mathbf{w}) \propto \exp\left( -\frac{1}{2} (\mathbf{w} - \hat{\mathbf{w}})^T \tilde{C}^{-1} (\mathbf{w} - \hat{\mathbf{w}}) \right),
\end{align}
where $\hat{\mathbf{w}}$ is the maximum-likelihood estimate of $\mathbf{w}$ and $\tilde{C}$ the covariance matrix in the space of the $\mathbf{w}$ parameters. 
Furthermore, the evidence corresponding to $\mathbf{D}_1$ can be computed analytically with the Laplace method 
(see for instance Ref.~\cite{10.5555/971143}) in the approximation of broad uniform priors:
\begin{align}
\ln \mathcal{Z}_1 = -\frac{1}{2} (\mathbf{D}_1 - \hat{\mathbf{D}}_1)^T C_1^{-1} (\mathbf{D}_1 - \hat{\mathbf{D}}_1) 
+ \ln \frac{\sqrt{(2\pi)^{N} |\tilde{C}|}}{\displaystyle \prod_{i=1}^{N} (b_i - a_i)},
\end{align}
where the second term is the \textit{Occam factor}, penalising unnecessarily complex models through their prior volume (note 
that $N$ denotes the number of parameters of the model, i.e. the number of weights being used).

The Gaussian posterior from $\mathbf{D}_1$ can be used as a \textit{prior} for the remaining fit to $\mathbf{D}_2$:
\begin{align}
p(\mathbf{w} | \mathbf{D}) \propto p(\mathbf{w} | \mathbf{D}_1)\, p(\mathbf{D}_2 | \mathbf{w}).
\end{align}
The final posterior is then sampled numerically using nested sampling and the final evidence is simply the sum of the contributions from the analytic and numerical stages:
\begin{align}
\ln \mathcal{Z} = \ln \mathcal{Z}_1 + \ln \mathcal{Z}_2.
\end{align}

Such Bayesian updating approach, described in more detail in App.~\ref{app:bayesian_update}, ensures both efficiency and consistency. 
We exploit analytic results where possible, and rely on robust numerical integration where necessary to account for non-linearities 
and theoretical constraints.

\subsection{Model selection and Bayesian model averaging}
\label{subsec:model_averaging}

The results of Sect.~\ref{sec:basis} demonstrate that our proposed linear models, based on the POD of the functional space of PDFs, 
can effectively span the PDF space, provided a sufficient number of basis elements are included. 
This raises a central question: how many basis elements should be used from our POD basis? While greater flexibility allows a model 
to describe more features, it also increases complexity. The optimal model is therefore the simplest one that provides an adequate 
description of the data, a realization of Occam’s razor.

Bayesian statistics offers a natural framework for addressing this balance through the Bayesian evidence, defined as:
\begin{align}
\mathcal{Z} = p(\vec{D} | \mathcal{M}) = \int d\vec{w}\, p(\vec{D}|\vec{w}) p(\vec{w}),
\end{align}
where $\mathcal{M}$ is the model (where in this work, the models $\mathcal{M}$ differ only in the number of basis elements which are included from the POD basis), $\vec{w}$ are the model's parameters, $p(\vec{D}|\vec{w})$ the likelihood, and $p(\vec{w})$ the prior. The evidence represents the probability of the data given the model. Applying Bayes’ theorem, the posterior probability of a model $\mathcal{M}$ given the data $\vec{D}$ is
\begin{align}
p(\mathcal{M} | \vec{D}) = \frac{p(\vec{D} | \mathcal{M}) p(\mathcal{M})}{p(\vec{D})},
\end{align}
where $p(\mathcal{M})$ is the prior probability assigned to the model. When comparing two models, $\mathcal{M}_1$ and $\mathcal{M}_2$, the ratio of their posterior probabilities is
\begin{align}
\frac{p(\mathcal{M}_1 | \vec{D})}{p(\mathcal{M}_2 | \vec{D})} = \mathcal{B}_{12} \frac{p(\mathcal{M}_1)}{p(\mathcal{M}_2)},
\end{align}
where $\mathcal{B}_{12} = \mathcal{Z}_1 / \mathcal{Z}_2$ is the Bayes factor. If both models are assigned equal prior probability, model selection is driven directly by the Bayes factor, which quantifies the relative support the data provide for each model.

A key advantage of this approach is that it automatically balances goodness-of-fit and model complexity. 
The evidence integral rewards models that describe the data well but penalises those that achieve this only by 
introducing unnecessary complexity. Parameters that do not significantly improve the likelihood contribute little to 
the evidence, as their effect is diluted by the large prior volume. This mechanism provides a principled safeguard against 
over-fitting without the need for ad hoc regularisation.

Furthermore, rather than selecting a single model, we apply Bayesian model averaging, which accounts for model uncertainty by combining results from all considered models. 
The posterior distribution for any quantity of interest $\Delta$ is then
\begin{align}
p(\Delta | \vec{D}) = \sum_{\mathcal{M}_n \in \mathcal{A}} p(\Delta | \vec{D}, \mathcal{M}_n)\, p(\mathcal{M}_n | \vec{D}),
\end{align}
where $\mathcal{A}$ is the set of models, and $p(\mathcal{M}_n | D)$ is the posterior model probability,
\begin{align}
p(\mathcal{M}_n | \vec{D}) = \frac{p(\vec{D} | \mathcal{M}_n)\, p(\mathcal{M}_n)}{\sum_{\mathcal{M}_l \in \mathcal{A}} p(\vec{D} | \mathcal{M}_l)\, p(\mathcal{M}_l)} 
= \frac{p(\vec{D} | \mathcal{M}_n)}{\sum_{\mathcal{M}_l \in \mathcal{A}} p(\vec{D} | \mathcal{M}_l)},
\end{align}
where in the last step we have assumed equal prior probabilities for all models.
For numerical stability, especially when the evidences differ widely, we compute
\begin{align}
p(\mathcal{M}_n | \vec{D}) = \frac{\exp(\ln Z_n - \ln Z_{\mathrm{ref}})}{\sum_{\mathcal{M}_l \in \mathcal{A}} \exp(\ln Z_l - \ln Z_{\mathrm{ref}})},
\label{eq:model_prob_exponents}
\end{align}
where $\ln Z_{\mathrm{ref}}$ is typically the log-evidence of the model with highest evidence.

In practice, this averaging procedure naturally emphasises the models best supported by the data while 
down-weighting overly complex or underperforming models. Models with a log-evidence more than 4 units 
below the best one contribute negligibly to the model average, so only those within a certain range of the top model need to be retained. 
The final result reflects both the data constraints and the model uncertainties, yielding robust predictions that are less sensitive to specific choices of parametrisation.


\subsection{PDF delivery}
\label{subsec:deliverable}

In this section we briefly describe the way in which a PDF set is delivered once the posterior distribution for the PDF parameters $\mathbf{w}$, 
Eq.~\eqref{eq:posterior} has been determined from the set of models that have been selected using the procedure discussed in 
Sect.~\ref{subsec:model_averaging}. 

The posterior distribution of the weights $\vec{w}$ fully determines the probability distribution in the space of PDFs. 
To represent this posterior, we generate $N_{\rm rep}$ samples from the weight posterior, capturing the essential features of the distribution:
\begin{equation}
    f^{(i)}(x,Q_0^2) = f_{\vec{w}^{\,(i)}}(x), \qquad {\rm with} \qquad \vec{w}^{\,(i)} \sim P(\vec{w}|\vec{D}),
\end{equation}
for $i = 1, \ldots, N_{\rm rep}$. From this ensemble of replicas, standard statistical estimators, such as the mean and covariance matrix, can be computed in PDF space and subsequently propagated to the space of physical observables. 
To accurately reproduce central values and marginal uncertainties of a quasi-Gaussian posterior distribution, $N_{\rm rep} = \mathcal{O}(100)$ is typically sufficient. 
However, a larger number of replicas is required to reliably capture correlations at the percent level.

The $N_{\rm rep}$ posterior samples of the weights are converted into $N_{\rm rep}$ PDF replicas, which are made available through the standard LHAPDF interface~\cite{Buckley:2014ana}, similarly to the Monte Carlo replicas employed in conventional PDF fits.

\section{Validation of the methodology and uncertainty quantification}
\label{sec:results}
In this section, we discuss the results obtained using the linear parametrisation, in the context of a multi-closure test.
The section is structured as follows: in Sect.~\ref{subsec:data_description} we discuss the dataset included in the present analysis; 
in Sect.~\ref{subsec:validation_model_selection} we test the methodology and showcase how the adopted Bayesian 
model-selection methodology is successful in selecting the correct level of complexity. 
Finally, in Sect.~\ref{subsec:uq_data_region} we present the results of the multi-closure test and 
discuss the faithfulness of the PDF uncertainties yield by our methodology. 


\subsection{Data and closure test settings}
\label{subsec:data_description}
\begin{figure}
    \centering
    \includegraphics[width=0.8\linewidth]{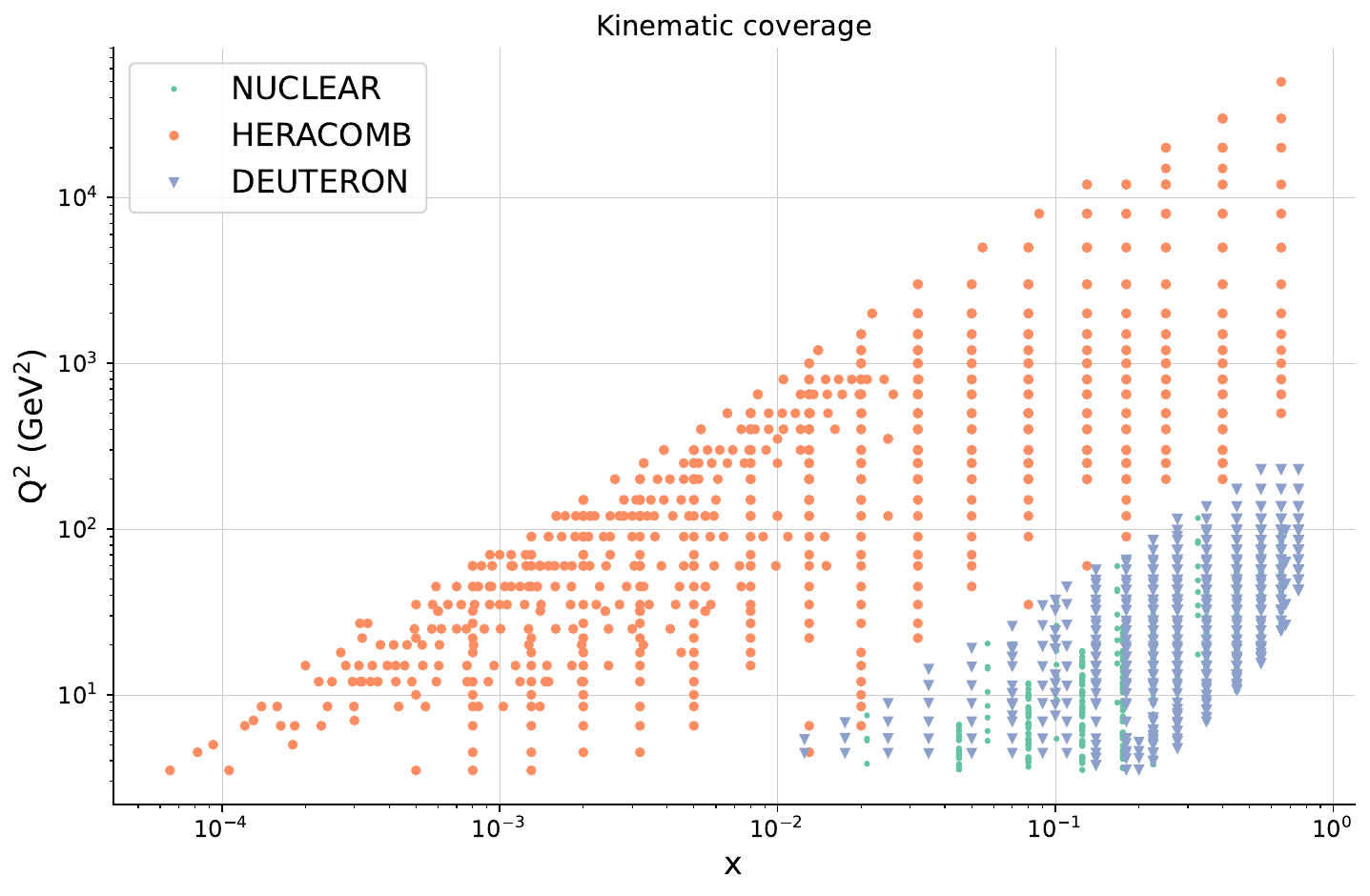}
    \caption{Kinematical coverage of DIS data included in the multi-closure test analysis presented in Sect.~\ref{sec:results}.}
    \label{fig:kin_coverage_dis}
\end{figure}

In this analysis, we consider the full Deep Inelastic Scattering (DIS) dataset employed in the NNPDF4.0 study~\cite{NNPDF:2021njg}. 
The kinematic coverage of the input dataset is shown in Fig.~\ref{fig:kin_coverage_dis} and a 
summary of the datasets used is provided in Table~\ref{tab:dis_data_table}. For each entry, we 
indicate the corresponding experiment; this distinction is important, as datasets from different experiments 
are uncorrelated, while those from the same experiment may exhibit internal correlations.
The DIS observables are modelled either:
\begin{itemize}
    \item \textit{linearly}, as a convolution of a partonic cross section with a single PDF; or
    \item \textit{as a ratio} of two cross sections (or structure functions), as in the case of the NMC data, which involve the ratio of $F_2$ deuteron over proton structure functions.
\end{itemize}

Taking advantage of known correlation structures, we partition the data into two experimentally uncorrelated subsets. 
The first subset includes data 
from the NMC measurements of the proton structure functions, which are uncorrelated to the ratio data, 
the fixed-target nuclear experiments CHORUS and NuTeV, and the HERA experiment. This subset comprises only linear observables, enabling a fast analytical fit. 
The second subset includes data involving deuterons, such as BCDMS, SLAC and the NMC dataset that measures the ratio between $F_2^d$ and $F_2^p$, 
which are all correlated either because of the experimental luminosity of because of deuteron nuclear corrections~\cite{Ball:2020xqw}. This subset requires a numerical fit using MCMC-like methods, as it introduces a non-linear dependence on the POD weights. As explained in Sect.~\ref{subsec:fitting_method}, this separation enhances computational 
efficiency: the analytical fit, being significantly faster, can be used to define a prior for the subsequent numerical fit.
Table~\ref{tab:dis_data_table} further distinguishes between the datasets used for PDF determination (in-sample) and those used to validate predictions (out-of-sample). 
Following a procedure similar to that of Ref.~\cite{Barontini:2025lnl}, the two samples are constructed to ensure that the in-sample and out-of-sample sets are 
statistically representative of the full dataset. 
%
\begin{table}[t!]
  \centering
  \small
  \begin{tabular}{l c c c}
    \toprule
    \textbf{Dataset} 
      & \textbf{\(N_{\text{data}}\)} 
      & \textbf{Experiment} 
      & \textbf{Model} \\
    \midrule
    \itshape Numerical fit \\
    \midrule
    NMC \(F_2^d/F_2^p\) \cite{Arneodo:1996kd} 
      &  121 & Deuteron 
      & Ratio      \\
    SLAC \(F_2^p\)  \cite{Whitlow:1991uw}   
      &   33 & Deuteron 
      & Linear   \\
    SLAC \(F_2^d\)  \cite{Whitlow:1991uw}   
      &   34 & Deuteron 
      & Linear     \\
    BCDMS \(F_2^p\) \cite{Benvenuti:1989rh} 
      &  333 & Deuteron 
      & Linear  \\
    BCDMS \(F_2^d\) \cite{Benvenuti:1989rh} 
      &  248 & Deuteron 
      & Linear  \\
    \midrule
    \multicolumn{4}{l} {\itshape Analytical fit} \\
    \midrule
    (*)  NMC \(\sigma^{\rm NC,p}\) \cite{Arneodo:1996qe}          
      &  204 & NMC       & Linear  \\
    CHORUS \(\sigma_{CC}^{\nu}\) \cite{Onengut:2005kv}         
      &  416 & Nuclear   & Linear  \\
    CHORUS \(\sigma_{CC}^{\bar\nu}\) \cite{Onengut:2005kv}     
      &  416 & Nuclear   & Linear  \\
    NuTeV dimuon \(\sigma_{CC}^{\nu}\) \cite{Goncharov:2001qe} 
      &   39 & Nuclear   & Linear  \\
    (*)  NuTeV dimuon \(\sigma_{CC}^{\bar\nu}\) \cite{Goncharov:2001qe} 
      &   37 & Nuclear   & Linear  \\
    HERA I+II \(\sigma_{CC}^{e^+p}\) \cite{Abramowicz:2015mha}  
      &   39 & HERA  & Linear  \\
    HERA I+II charm \(\sigma_{NC}\) \cite{H1:2018flt}           
      &   37 & HERA  & Linear  \\
    HERA I+II \(\sigma_{NC}^{e^-p}\) (320 GeV)                 
      &  159 & HERA  & Linear  \\
    (*)  HERA I+II \(\sigma_{NC}^{e^+p}\) (460 GeV)                 
      &  204 & HERA  & Linear  \\
    HERA I+II \(\sigma_{NC}^{e^+p}\) (575 GeV)                 
      &  254 & HERA  & Linear  \\
    HERA I+II \(\sigma_{NC}^{e^+p}\) (820 GeV)                 
      &   70 & HERA  & Linear  \\
    HERA I+II \(\sigma_{NC}^{e^+p}\) (920 GeV)                 
      &  377 & HERA  & Linear  \\
    (*)  HERA I+II \(\sigma_{CC}^{e^-p}\) \cite{Abramowicz:2015mha}  
      &   42 & HERA  & Linear  \\
    HERA I+II bottom \(\sigma_{NC}\) \cite{H1:2018flt}          
      &   26 & HERA  & Linear  \\
    \midrule
    \multicolumn{1}{l}{\textbf{Total}} 
      & 3089 &           &                  \\
    \bottomrule
  \end{tabular}
  \caption{Summary of DIS datasets included in the closure test analysis presented in Sect.~\ref{sec:results}.  
  Deuteron data (top) require a numerical MCMC fit due to the ratio observable, while NMC, nuclear target experiments and HERA data (bottom) 
  admit an analytical solution. The out-of-sample datasets are denoted by an asterisk.}
  \label{tab:dis_data_table}
\end{table}
%

%

\subsection{Validation of the model selection strategy}
\label{subsec:validation_model_selection}

The reliability and performance of our fitting and model selection methodology 
are assessed in the context of multi-closure test, by performing a series of fits on synthetic datasets that exhibit realistic, 
experiment-like features. This procedure follows the closure-test strategy employed by the NNPDF collaboration in studies such as those 
discussed in Refs.~\cite{Barontini:2025lnl,DelDebbio:2021whr}.

We begin by constructing an underlying ``true'' PDF from the POD basis 
\begin{equation}
[\boldsymbol\varphi_0, \boldsymbol{\varphi}_1, \dots, \boldsymbol{\varphi}_N],
\end{equation}
and fixing the weights to those of a model with dimension \(N = 40\):
\begin{equation}
  \mathbf{f}_{\rm in}
    = \boldsymbol\varphi_0
    + \sum_{k=1}^{N} w_k^{\rm in} \left(\boldsymbol{\varphi}_k - \boldsymbol\varphi_0\right).
\end{equation}
This PDF serves as the generative model for synthetic data. There are two such levels of data, namely 
\begin{equation}
  \mathbf{D}^{L_0} 
    = T[\mathbf{f}_{\rm in}],
    \label{eq:L0data}
\end{equation}
which is dubbed as Level-0 data, and is nothing but the underlying law itself, built by convolving \(\mathbf{f}_{\rm in}\) with partonic cross sections.  
Adding Gaussian noise generated from the covariance matrix of the input data we obtain what is called Level-1 data, namely:
\begin{equation}
  \mathbf{D}^{L_1} 
    = T[\mathbf{f}_{\rm in}] + \boldsymbol\eta,
    \label{eq:L1data}
\end{equation}
where \(\boldsymbol\eta \sim \mathcal{N}(\mathbf{0}, C)\), and \(C\) is the covariance matrix used in the fit.
The likelihood function used in the closure test is given in Eq.~\eqref{eq:general_likelihood}, with the 
Lagrangian penalty coefficients for the positivity and integrability constraints set to $\Lambda_{\rm pos}=100$ and $\Lambda_{\rm int}=50$  
respectively. We have verified that increasing both penalty terms by up to a factor of 3 produces results that are statistically consistent.
Note that in this setup all other hyperparameters, such as the preprocessing exponents, are kept fixed, so the likelihood depends solely on the inference parameters, namely the POD weights. Consequently, when we refer to the model complexity or its parametric dimension in what follows, we refer exclusively to the dimensionality of the POD representation, that is, the number of POD weights.

To validate our model-selection strategy, we perform a scan over a family of models with progressively 
increasing complexity, each fitted to the same instance of Level-1 data. By comparing the Bayesian evidence 
for each model, we verify that the inference procedure correctly identifies the model with the appropriate 
number of parameters, thus preventing both under-fitting and over-fitting.

\begin{figure}[t!]
    \centering
    \includegraphics[width=0.7\linewidth]{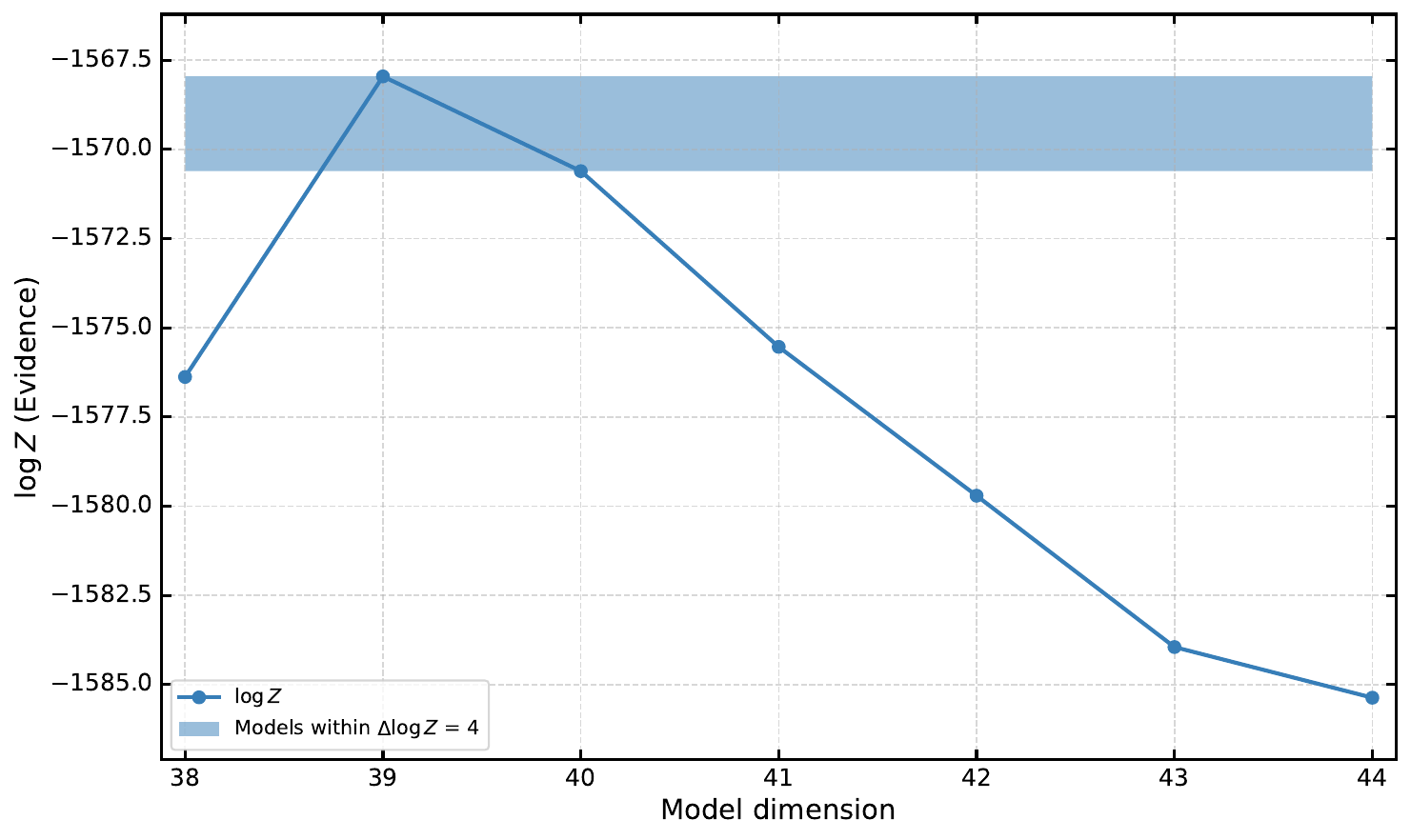}
    \caption{Bayesian evidence for models with increasing complexity, fitted to a Level-1 dataset, Eq.~\eqref{eq:L1data}, 
    generated using the same random seed. The model dimension indicates the number of basis function that we include in the parametrisation.}
    \label{fig:logz_l1_scan}
\end{figure}

In Fig.~\ref{fig:logz_l1_scan}, we plot the total evidence as a function of model complexity. 
We observe that the model with 39 parameters is favoured, closely followed by the one with 40 parameters.
The best-fit \(\chi^2/N_{\rm data}\) and the model posterior probabilities \(p(\mathcal{M}_k | D)\), computed 
using Eq.~\eqref{eq:model_prob_exponents} and considering only the two most likely 
models \(\mathcal{A} = \{\mathcal{M}_{39}, \mathcal{M}_{40}\}\), are reported in Table~\ref{tab:bayes_factor_chi2_tab}.

\begin{table}[htb!]
    \centering
    \renewcommand{\arraystretch}{1.2}
    \caption{Goodness-of-fit and model posterior probabilities for the two most likely models.}
    \label{tab:bayes_factor_chi2_tab}
    \begin{tabular}{lcc}
        \hline
        \textbf{Model Complexity} & \(\chi^2 / N_{\rm data}\) & \(p(\mathcal{M}_k | D)\) \\
        \hline
        39 & 1.033 & 0.934 \\
        40 & 1.032 & 0.066 \\
        \hline
    \end{tabular}
\end{table}

Interestingly, the underlying law used to generate the data includes 40 non-zero components, making it in principle compatible with the 40-parameter model. However, as shown in Table~\ref{tab:bayes_factor_chi2_tab}, the improvement in fit quality for the more complex model is marginal. As a result, the simpler 39-parameter model is preferred, since 
the additional parameter in the 40-parameter model is not sufficiently constrained by the data and is therefore penalised by the Occam factor. 

It is important to emphasise that this outcome does not indicate any flaws in the inference procedure. On the contrary, it highlights a fundamental feature of 
Bayesian model comparison: when the data lack sufficient sensitivity to constrain an additional parameter, the simpler model is naturally preferred. 
This behaviour is not only expected but also desirable, as it guards against overfitting and unwarranted model complexity. With more precise or additional data, 
sensitivity to the 40\textsuperscript{th} parameter may well emerge, at which point the data themselves will justify the inclusion of that parameter.

\begin{figure}[t!]
    \centering
    \begin{subfigure}[t]{0.48\textwidth}
        \centering
        \includegraphics[width=\linewidth]{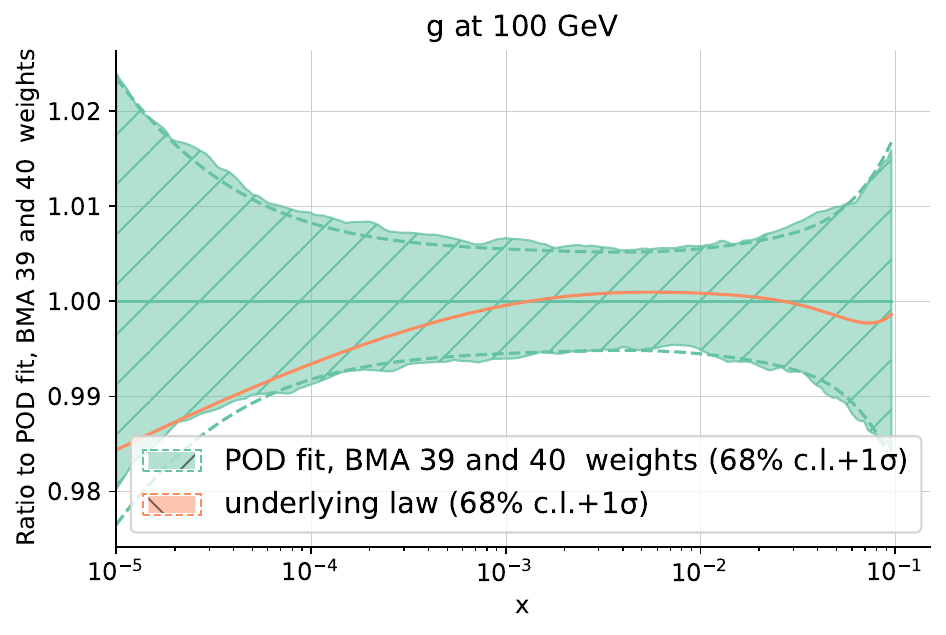}
    \end{subfigure}
    \hfill
    \begin{subfigure}[t]{0.48\textwidth}
        \centering
        \includegraphics[width=\linewidth]{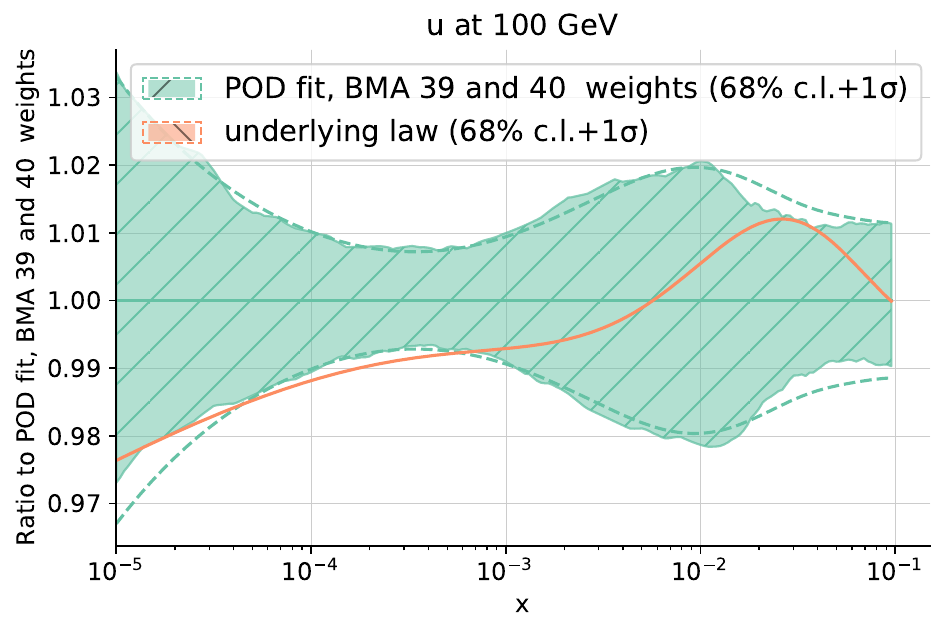}
    \end{subfigure}


    \begin{subfigure}[t]{0.48\textwidth}
        \centering
        \includegraphics[width=\linewidth]{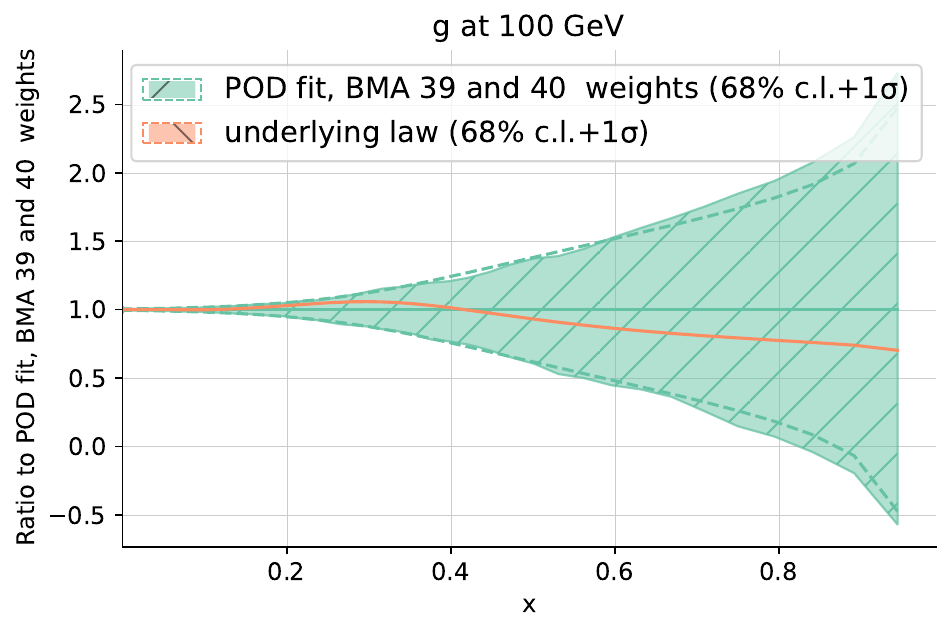}
    \end{subfigure}
    \hfill
    \begin{subfigure}[t]{0.48\textwidth}
        \centering
        \includegraphics[width=\linewidth]{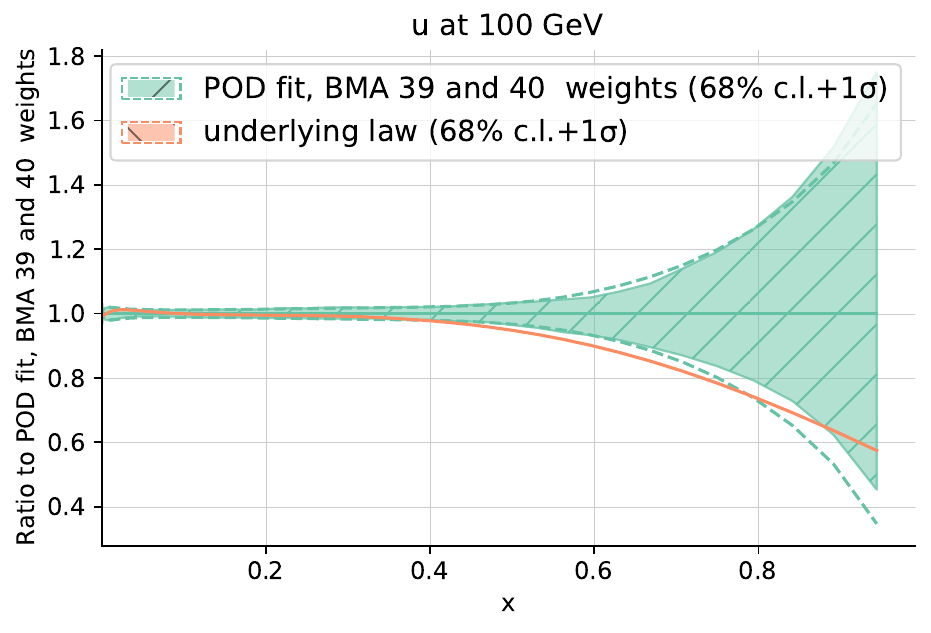}
    \end{subfigure}

    \caption{Model averaged gluon and up-quark valence PDFs with $68\%$ uncertainty bands against the underlying law both in logarithmic as well as linear scale.}
    \label{fig:2x2grid_gluon_and_up}
\end{figure}

For completeness, in Fig.~\ref{fig:2x2grid_gluon_and_up} we present the model-averaged gluon and up-quark valence PDFs, together with their corresponding $68\%$ uncertainty bands, compared against the underlying law used to generate the synthetic data. The excellent agreement demonstrates that the fit successfully recovers the true underlying PDFs within the estimated uncertainties.




\subsection{Uncertainty quantification}
\label{subsec:uq_data_region}
To validate the reliability of PDF-predicted uncertainties in data space, we employ the normalised bias estimator recently introduced in Ref.~\cite{Barontini:2025lnl}. 
This metric quantifies the actual mean square deviation of predictions from ground truth, expressed in units of the predicted standard deviation.
We perform $N_{\rm fit} = 25$ closure tests\footnote{Previous studies have demonstrated that this sample size provides enough statistics
to obtain reliable results.}, each utilizing a distinct, randomly generated set of $L_1$ data, indexed by $l$. For each fit $l$, we compute the 
normalised bias
\begin{align}
    \label{eq:bias_definition}
        B^{(l)} & = \frac{1}{N_{\rm data}}\sum_{i,j=1}^{N_{\rm data}} \left(\mathbb{E}_{\epsilon}T_i[f_{k}^{(l)}]- D^{L_0}_i\right) (C_{\rm PDF})_{ij}^{-1} \left(\mathbb{E}_{\epsilon}T_j[f_{k}^{(l)}]- D^{L_0}_j\right) ,
\end{align}
where the expectation $\mathbb{E}_{\epsilon}$ averages over the posterior sample in data space: 
\begin{align}
    \mathbb{E}_{\epsilon}T_i[f^{(l)}] = \frac{1}{N_{\rm rep}} \sum_{k=1}^{N_{\rm rep}} T_i[f_{k}^{(l)}] .
\end{align}
Here $N_{\rm rep}$ is the sample size used to approximate the posterior distribution of the $N$-dimensional $\bf{w}$ parameter space, 
as discussed in Sect.~\ref{subsec:deliverable}. $C_{\rm PDF}$ denotes the PDF covariance matrix derived from the Monte Carlo replicas 
sampling the posterior distribution, and $D^{L_0}_i$ is the theoretical prediction computed from the 
known underlying law, Eq.~\eqref{eq:L0data}.
After computing the average over the $N_{\rm fit}$ and taking the square root
\begin{align}
  \label{eq:bias_variance_ratio_definition}
    R_{b} = \sqrt{\frac{1}{N_{\rm fit}}\sum_{l=1}^{N_{\rm fit}} B^{(l)} } ,
\end{align}
we estimate the uncertainties associated with finite sample sizes using a bootstrap procedure, following the methodology 
detailed in Appendix B of Ref.~\cite{Barontini:2025lnl}. 

  %
\begin{figure}[tb]
    \centering
    \includegraphics[width=0.75\linewidth]{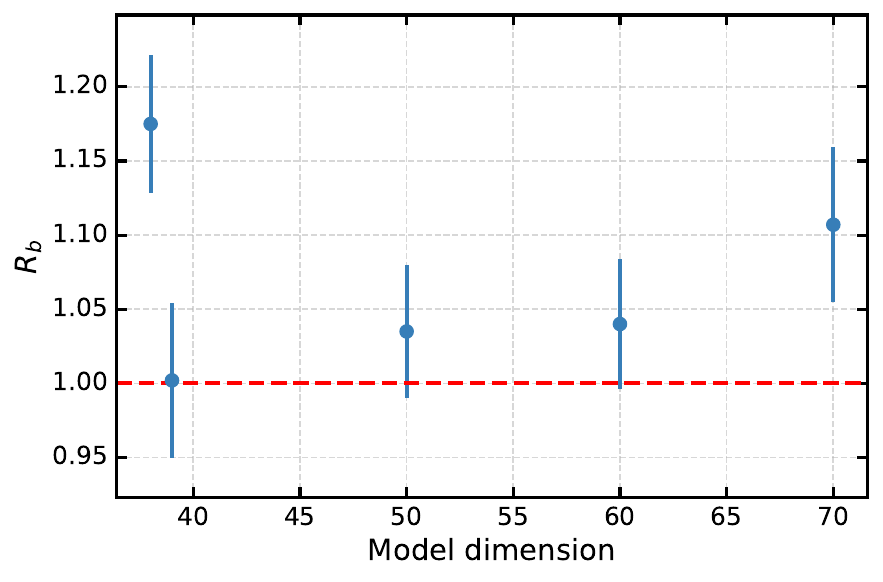}
    \caption{Normalised bias, Eq. \eqref{eq:bias_variance_ratio_definition}, computed on the out-of-sample dataset for several models, with complexities of 38, 39, 50, 60 and 70.}
    \label{fig:norm_bias_vs_model_complexity}
\end{figure}
In Fig.~\ref{fig:norm_bias_vs_model_complexity} we present the normalised bias computed on the out-of-sample data subset as a 
function of model complexity.
The results reveal that both under-parametrised and over-parametrised models (as compared to the underlying law) systematically underestimate uncertainties 
in the data region.
%
This finding highlights the fact that when the correct underlying model is obtained, then the Bayesian analysis perfectly reproduces the statistical distribution of the data that entered the fit. On the other hand, when under-parametrised or over-parametrised models are used, we see that the fit does not reproduce the statistical distribution of the data to such a faithful extent. 
The method we have proposed allows for correct identification of the model complexity, up to an averaging across a combination of potential candidate models, accounting for both the possibility of over- and under-parametrisation. 

\section{Conclusions}
\label{sec:conclusion}

In this work, we have proposed and tested a new efficient Bayesian framework for PDF fits, 
based on a novel linear parametrisation of PDF space. This work provides the foundation for future, realistic fits using this framework.

In Sect.~\ref{sec:methodology}, we introduced the method of \textit{proper orthogonal decomposition} (POD), 
central to our workflow. This method allows for the approximate reduction of a large, potentially non-linear, 
space to a linear span of basis functions. The basis functions are arranged in order of importance, 
allowing for flexible precision. They also inherit any homogeneous, linear properties that are characteristic 
of the original space; this is important because it ensures that for a PDF space, the candidate basis functions obey integrability conditions and sum rules.

In Sect.~\ref{sec:basis}, we used POD to reduce a candidate space of PDFs in terms of a small set of basis functions. 
The space we began with was a space of extremely flexible functions, given by randomly initialised neural network. 
The use of such a space allows the approximation of a large variety of functional forms. 
We explicitly demonstrated that the linear reduction of the neural network space, via POD, was 
able to approximate both the original neural network space, and also other functional 
forms well (that is, the reduction retained the flexibility of the original neural network space).

In Sect.~\ref{sec:model-selection} we discussed how to use the basis we obtained via POD of 
the neural network space in a fully Bayesian fit of PDFs. After 
introducing the likelihood function and the penalty terms, relating to positivity and integrability that are used in the fit, we  
described how the fit can be accelerated using a Bayesian updating strategy; 
this strategy demonstrates the extremely advantageous feature  of the Bayesian methodology, and showcases the idea 
of `updating knowledge based on data'. Further, we described the powerful 
method of \textit{Bayesian model selection} and \textit{model averaging}, which allows for a dynamical choice of 
the number of basis functions from our linear parametrisation of PDFs that are required in a given fit. 
Importantly, this simultaneously guards against under-fitting and over-fitting in the Bayesian setting.
The delivery of the Monte Carlo PDF set obtained by sampling the posterior probability is also discussed. 

Finally, in Sect.~\ref{sec:results}, we presented the main results of our work. In the context of a multi-closure test, we applied our framework 
to a fully Bayesian fit of synthetic DIS data, including both data where the PDFs enter 
linearly and data where the PDFs enter non-linearly (in certain ratio observables). Applying Bayesian model selection and averaging, 
we both accurately predicted the correct underlying model complexity using our Bayesian workflow, and demonstrated excellent recovery 
of the underlying physical law used to generate the data. We also performed a rigorous investigation of the uncertainties produced 
using our method and demonstrated that, through the use of Bayesian model selection, the difference between the normalised bias and unity 
is minimised in a multi-closure test when our workflow is applied, showcasing the excellent faithfulness of the PDF uncertainties produced using this method.

In conclusion, this study presents a novel, flexible, efficient representation of PDF space that can be used seamlessly with a 
Bayesian framework, making full use of the inference tools that are provided to us through Bayesian statistics, 
such as model selection and averaging. 
The workflow in this study was rigorously tested in a closure-test setting, paving the way for a future Bayesian fit of the global dataset in subsequent works. 

Motivated by the principles of Open and FAIR~\cite{FAIR} (findable, accessible, interoperable and reusable) Science, 
and to promote open-source science, we have made the code used to produce the results of this paper available from the {\sc\small GitHub} repository:
\begin{center}
  {\tt \url{https://github.com/HEP-PBSP/wmin-model}}.
\end{center}
The code is accompanied by documentation and tutorials at:
\begin{center}
 {\tt \url{https://github.com/comane/NNPOD-wiki}}.
\end{center}
Development of this code was made possible by the following open-source projects: \cite{zaharid_2025_14803041,NNPDF:2021uiq,candido_2025_15655649}.  
Future developments, including the production of PDF fits based on real datasets, including also hadronic observables, will 
rely on the availability of the general-purpose, flexible, fast PDF-fitting platform, \texttt{colibri}~\cite{Costantini:2025agd}.

\section*{Acknowledgments}
We are grateful to Shayan Iranipour and Zahari Kassabov for their work in the initial ideas and 
for their work on an earlier implementation of the project.
We are extremely grateful to the members of the NNPDF collaboration for their suggestions regarding
this work; in particular, we are indebted to Juan Cruz Martinez, Luigi del Debbio and Tommaso Giani for useful discussions during the development of this work.
M.~N.~C., L.~M. and M.~U. are supported by the European Research Council under the
European Union's Horizon 2020 research and innovation Programme (grant agreement n.950246).
L.~M. acknowledges support from the European Union under the MSCA fellowship (Grant agreement N. 101149078) {\it Advancing global SMEFT fits in the LHC precision era (EFT4ward)}.
They are also  partially supported by the STFC grant ST/T000694/1. J.~M.~M. is supported by the 
donation of Christina and Peter Dawson to Lucy Cavendish College.

\appendix
\section{Proper orthogonal decomposition: further detail}
\label{app-infinite_dimensional_pod}
In this Appendix we provide more details about the POD discussed in Sect.~\ref{subsec:pod}, showing that the POD procedure (in the finite-dimensional case) can be reduced to 
a problem of singular value decomposition (SVD). Further discussion is given in Sect.~3.4.2 of Ref.~\cite{Holmes_Lumley_Berkooz_Rowley_2012}. We also discuss the implementation of theory constraints on the POD basis.

\paragraph{Proof of equivalence of POD and SVD.} To see the equivalence of POD and singular value decomposition, observe that we can combine the initial ensemble into a \textit{data matrix}:
\begin{align}
    \tilde{X} = \bigg[\tilde{\mathbf{g}}_1,\ldots, \tilde{\mathbf{g}}_M\bigg].
\end{align}
Let $\tilde{X} = U\Sigma V^T$ be the singular value decomposition of this matrix, with $U = [\pmb{\varphi}_1, ..., \pmb{\varphi}_N]$, $\Sigma$ is block-diagonal, with entries $\sigma_1, ..., \sigma_N$ along the leading diagonal, and $V = [\vec{v}_1, ..., \vec{v}_M]$. Then we can write:
\begin{equation}
    \tilde{X} = \sum_{k=1}^{r} \sigma_k \pmb{\varphi}_k \vec{v}_k^T,
\end{equation}
where $r$ is the rank of $\tilde{X}$. Now, observe that the eigenvalue problem for the autocorrelation matrix $A \vec{e} = \lambda \vec{e}$ can equivalently be written as:
\begin{equation}
    \tilde{X} \tilde{X}^T \vec{e} = \lambda \vec{e}.
\end{equation}
We can now check that $\pmb{\varphi}_k$ are the solutions of this eigenvalue problem. We have:
\begin{equation}
    \tilde{X}\tilde{X}^T \pmb{\varphi}_k = \sum_{i=1}^{r} \sum_{j=1}^{r} \sigma_i \sigma_j \pmb{\varphi}_i \vec{v}_i^T \vec{v}_j \pmb{\varphi}_j^T \pmb{\varphi}_k = \sum_{i=1}^{r} \sum_{j=1}^{r} \sigma_i \sigma_j \pmb{\varphi}_i \delta_{ij} \delta_{jk} = \sigma_k^2 \pmb{\varphi}_k.
\end{equation}
Hence, the columns of $U$ are the solutions of the POD problem; thus, we can obtain an easy solution to POD in the finite-dimensional setting using SVD.

In practice, we choose the POD basis vectors as the columns of $U \Sigma$, and then form our basis matrix $B$ by adding the mean vector $\varphi_0$ to each column:
\begin{align}
    B = U \Sigma + \boldsymbol{\varphi}_0 = \bigg[\mathbf{b}_1,\ldots, \mathbf{b}_r\bigg] .
    \label{eq:pca_basis_elements}
\end{align}
Given the basis vectors defined in Eq. \eqref{eq:pca_basis_elements} we build our linear model as the standard LHAPDF interpolation~\cite{Buckley:2014ana} of the vectors:
\begin{align}
    \vec{f}_{\vec{w}} = \pmb{\varphi}_0 + \sum_{i=1}^{r} w_i \left(\vec{b}_i - \pmb{\varphi}_0 \right) , 
\end{align}
where the vectors consist of the relevant PDFs evaluated on the grid. Note that the precise grid used in practice is the LHAPDF grid, but in the fit the fast kernel table grid may not match this - therefore, it is important to state that we use the standard LHAPDF interpolation for our model for points in between the LHAPDF grid points.

\paragraph{Theory constraints on the POD basis.} It is a general property of POD that any linear, homogeneous property of the original sample is preserved in the resulting basis vectors; a full proof of this result can be found in Ref.~\cite{Holmes_Lumley_Berkooz_Rowley_2012}. Hence, by construction, the basis vectors $\varphi_k(x)$ necessarily satisfy the theory conditions we outlined in the previous section. 

To see why this holds in a simplified scenario, let us consider the sum rules for the PDFs. In the discrete case, the sum rules can be viewed as the discrete integral of the columns of the centred data matrix $\tilde{X}$.\footnote{Note that this holds directly for a single flavour in the evolution basis; the generalised flavour-stacked case reduces to a superposition of single-flavour cases.} Assuming an equally spaced discretised $x$-grid, the centred data matrix $\tilde{X}$ satisfies
\[
  \mathbf{1}_N^\top \tilde{X} \;=\; \mathbf{0}_M,
\]
where \(\mathbf{1}_N \in \mathbb{R}^N\) is the vector of ones and \(\mathbf{0}_M \in \mathbb{R}^M\) is the zero vector. Writing $\tilde{X}$ in terms of its singular value decomposition, $\tilde{X} \;=\; U \Sigma V^\top$, as we did above, we then have:
\[
  \mathbf{1}_N^\top (U \Sigma)
  \;=\; (\mathbf{1}_N^\top \tilde{X})\,V
  \;=\; \mathbf{0}_M^\top V
  \;=\; \mathbf{0}_N^\top.
\]
Therefore, if the central basis vector \(\boldsymbol{\varphi}_0\) satisfies the sum rule, then each column of the basis matrix defined in Eq. \eqref{eq:pca_basis_elements}
also satisfies the same sum rules. This precisely matches the discussion of the theory constraints in Section~\ref{sec:methodology}.
\section{A trick for the efficient computation of the parameters posterior}
\label{app:bayesian_update}
\noindent In the case of multivariate normal data consisting of several uncorrelated measurements, the posterior
distribution of the model parameters used to describe the data can be computed iteratively as we show below. This
can potentially help reduce the computational time needed for the evaluation of the posterior in particular
when new, uncorrelated, data is included.

\paragraph{The method}\mbox{}\\
\noindent Let us suppose that experimental data comprising $N_{\rm dat}$ datapoints is distributed according to a multivariate normal distribution
\begin{equation}
\vec{D} \sim N(\vec{t}(\vec{c}), C),
\end{equation}
where $C$ is an $N_{\rm dat}\times N_{\rm dat}$ experimental covariance matrix, and $\vec{t}:\mathbb{R}^{N_{\rm param}}\to \mathbb{R}^{N_{\rm dat}}$ is a smooth theory
prediction function (forward map), taking as argument a vector of $N_{\rm param}$ unknown theory parameters $\vec{c}\in \mathbb{R}^{N_{\rm param}}$.
Since in Bayesian statistics, $\vec{c}$ itself is assumed to be a random variable, it has some associated prior probability density $\pi(\vec{c})$ which in the following we
will assume to be a \quotes{sufficiently} wide uniform probability density.
Bayes' theorem then tells us that after an observation $\vec{D}_0$ of $\vec{D}$, the probability density of $\vec{c}$ is
\begin{equation}
\label{eq: bayes theorem}
p(\vec{c}|\vec{D}_0) = \frac{\pi{(\vec{c})} L(\vec{D}_0|\vec{c})}{Z} = \frac{\pi{(\vec{c})} \exp(-\frac{1}{2}||\vec{D}_0 - \vec{t}(\vec{c})||^2_{C}) }{\int d\vec{c} \; \pi{(\vec{c})} \exp(-\frac{1}{2}||\vec{D}_0 - \vec{t}(\vec{c})||^2_{C})},
\end{equation}
where we wrote the generalised $L_2$ norm as $||\vec{x}||^2_{C} = \vec{x}^TC^{-1}\vec{x}$ for a vector $\vec{x} \in \mathbb{R}^{N_{\rm dat}}$.\\
Now let's assume that $\vec{D}_0 = (\vec{D}_1, \vec{D}_2)^T$ with $\vec{D}_1\in \mathbb{R}^{n_1}$, $\vec{D}_2\in \mathbb{R}^{n_2}$ and $n_1+n_2=N_{\rm dat}$, and that the 
two measurements are uncorrelated, that is we can write the covariance matrix as $C = C_1 \bigoplus C_2$ with $C_1 \in \mathbb{R}^{n_1\times n_1}$
and $C_2 \in \mathbb{R}^{n_2\times n_2}$.
In this case, since the likelihood $L(\vec{D}_0|\vec{c})$ factorises ($C$ is block diagonal), we can write Eq. (\ref{eq: bayes theorem}) as
\begin{equation}
\label{eq: bayes for uncorrelated meas}
p(\vec{c}|\vec{D}_0) = \frac{\pi{(\vec{c})} \exp(-\frac{1}{2}||\vec{D}_1 - \vec{t}_1(\vec{c})||^2_{C_1}) \exp(-\frac{1}{2}||\vec{D}_2 - \vec{t}_2(\vec{c})||^2_{C_2})}{\int d\vec{c} \; \pi{(\vec{c})} \exp(-\frac{1}{2}||\vec{D}_1 - \vec{t}_1(\vec{c})||^2_{C_1}) \exp(-\frac{1}{2}||\vec{D}_2 - \vec{t}_2(\vec{c})||^2_{C_2})},
\end{equation}
where we wrote $\vec{t}(\vec{c}) = (\vec{t}_1(\vec{c}), \vec{t}_2(\vec{c}))^T$.\\
Now, by noticing that the posterior distribution of the parameters $\vec{c}$ in the case in which the observed data is given by $\vec{D}_1$ is given by
\begin{equation}
\label{eq: pd1 post}
p_{\vec{D}_1}(\vec{c}|\vec{D}_1) = \frac{\pi{(\vec{c})} \exp(-\frac{1}{2}||\vec{D}_1 - \vec{t}_1(\vec{c})||^2_{C_1})}{\int d\vec{c} \; \pi{(\vec{c})} \exp(-\frac{1}{2}||\vec{D}_1 - \vec{t}_1(\vec{c})||^2_{C_1})} = \frac{\pi{(\vec{c})} \exp(-\frac{1}{2}||\vec{D}_1 - \vec{t}_1(\vec{c})||^2_{C_1})}{Z_1} ,
\end{equation}
we can rewrite Eq. (\ref{eq: bayes for uncorrelated meas}) as
\begin{equation}
p(\vec{c}|\vec{D}_0) = \frac{ (Z_1 p_{\vec{D}_1}(\vec{c}|\vec{D}_1) ) \exp(-\frac{1}{2}||\vec{D}_2 - \vec{t}_2(\vec{c})||^2_{C_2})}{\int d\vec{c} \; (Z_1 p_{\vec{D}_1}(\vec{c}|\vec{D}_1) ) \exp(-\frac{1}{2}||\vec{D}_2 - \vec{t}_2(\vec{c})||^2_{C_2})} = \frac{ p_{\vec{D}_1}(\vec{c}|\vec{D}_1) \exp(-\frac{1}{2}||\vec{D}_2 - \vec{t}_2(\vec{c})||^2_{C_2})}{\int d\vec{c} \; p_{\vec{D}_1}(\vec{c}|\vec{D}_1) \exp(-\frac{1}{2}||\vec{D}_2 - \vec{t}_2(\vec{c})||^2_{C_2})} .
\end{equation}
Note that if we have a measurement $\vec{D} \sim N(\vec{t}(\vec{c}), C)$ such that $C = C_1 \bigoplus C_2 \bigoplus \ldots \bigoplus C_n$, 
we can apply this trick recursively and compute the posterior for $\vec{c}$ given the observed data $\vec{D}_0$ as
\begin{equation}
p(\vec{c}|\vec{D}_0) = \frac{\prod_{i=1}^{n-1} p_{\vec{D}_i}(\vec{c}|\vec{D}_i) \exp(-\frac{1}{2}||\vec{D}_n - \vec{t}_n(\vec{c})||^2_{C_n})}{\int d\vec{c} \; \prod_{i=1}^{n-1} p_{\vec{D}_i}(\vec{c}|\vec{D}_i) \exp(-\frac{1}{2}||\vec{D}_n - \vec{t}_n(\vec{c})||^2_{C_n})},
\end{equation}
with $p_{\vec{D}_1}(\vec{c}|\vec{D}_1)$ as given in Eq. (\ref{eq: pd1 post}) and $p_{\vec{D}_k}(\vec{c}|\vec{D}_k) = \frac{\prod_{i=1}^{k-1} p_{\vec{D}_i}(\vec{c}|\vec{D}_i) \exp(-\frac{1}{2}||\vec{D}_k - \vec{t}_k(\vec{c})||^2_{C_k})}{\int d\vec{c} \; \prod_{i=1}^{k-1} p_{\vec{D}_i}(\vec{c}|\vec{D}_i) \exp(-\frac{1}{2}||\vec{D}_k - \vec{t}_k(\vec{c})||^2_{C_k})}$ for $k > 1$.

\begin{figure}[t!]
\centering
\includegraphics[scale=0.5]{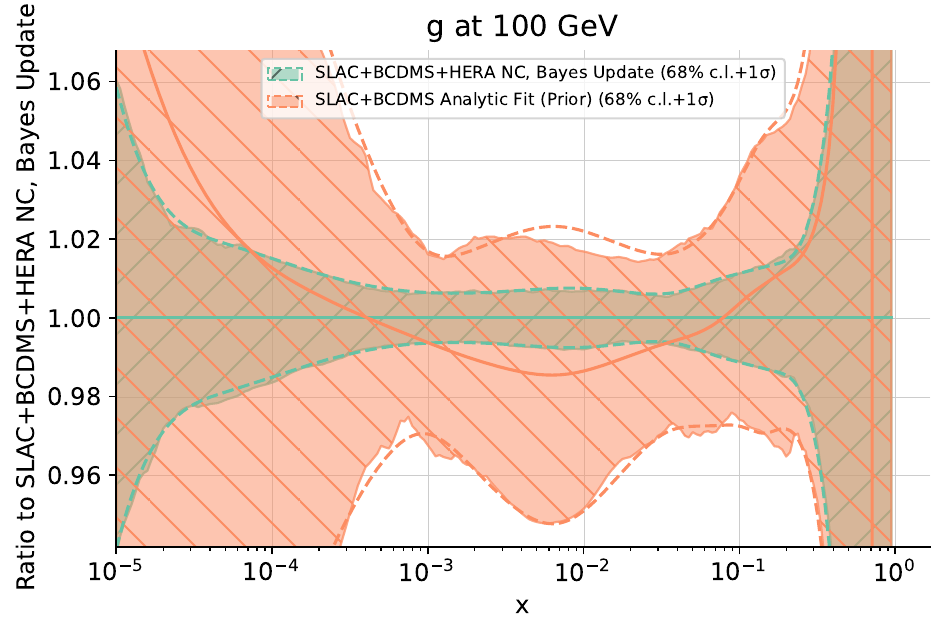}
\caption{In orange the \quotes{prior} PDF obtained by fitting SLAC + BCDMS data only. In green the posterior PDF obtained by fitting HERA NC data 
using as a prior the SLAC + BCDMS only fit.}
\label{fig:analytic_prior_vs_posterior_appendix}
\end{figure}
\begin{figure}[t!]
\centering
\includegraphics[scale=0.5]{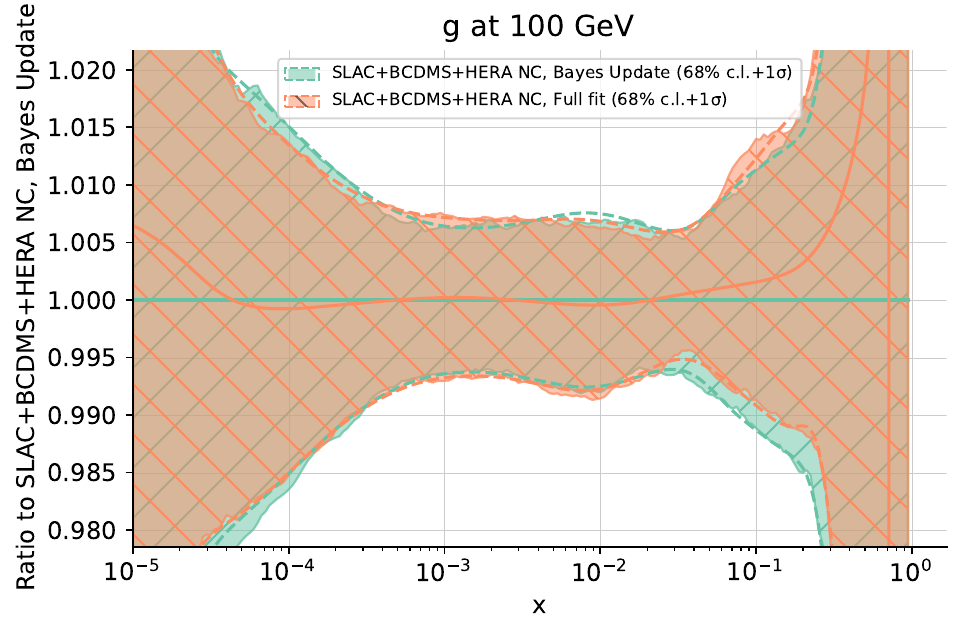}
\caption{Comparison between PDFs fitted on SLAC+BCDMS+HERA NC data, in orange using a uniform prior and fitting the data from the three experiments, in green using the SLAC+BCDMS prior and fitting data from the HERA NC experiment only.}
\label{fig:bayesian_updated_vs_fit_appendix}
\end{figure}
\paragraph{An explicit example}\mbox{}\\
To demonstrate the above method we apply it here for a simplified example. We consider data from three uncorrelated DIS experiments, namely, SLAC \cite{Whitlow:1991uw}, BCDMS \cite{BCDMS:1989qop}, and HERA NC \cite{H1:2015ubc}. We first compute a fit on the SLAC + BCDMS data and then use this fit as the prior distribution from which to sample from to compute the posterior distribution on the PDF parameters given all the data from the three experiments.
In Figure (\ref{fig:analytic_prior_vs_posterior_appendix}) we plot the \quotes{prior} PDF computed on the SLAC + BCDMS data versus the posterior PDF obtained by fitting HERA NC data only and using as a prior the SLAC + BCDMS only PDF.
In Figure (\ref{fig:bayesian_updated_vs_fit_appendix})
we display the agreement between the posterior PDF distributions obtained in green using the Bayesian updating method and in orange by fitting directly all the data using a uniform prior.

\bibliographystyle{utphys}
\bibliography{references}

\end{document}